\documentclass[10pt,journal,compsoc]{IEEEtran}
\ifCLASSOPTIONcompsoc
  \usepackage[nocompress]{cite}
\else
  \usepackage{cite}
\fi
\ifCLASSINFOpdf
\else
\fi
\usepackage{amsmath,amssymb,amsfonts}
\newtheorem{thm}{Theorem}

\newtheorem{defn}[thm]{Definition} 

\usepackage{array}
\usepackage[ruled,vlined]{algorithm2e}
\usepackage{algorithmic}
\usepackage{graphicx}
\usepackage{subfigure} 
\usepackage{textcomp}
\usepackage{color} 
\usepackage{soul}
\usepackage{multirow}
\usepackage{amsfonts}
\usepackage{enumerate}
\definecolor{blue}{RGB}{0,0,0}
\definecolor{red}{RGB}{223,58,45}

\begin{document}
%
\title{Achieving Fair-Effective Communications and Robustness in Underwater Acoustic Sensor Networks: A Semi-Cooperative Approach}
%
%
%
%

\author{Yu Gou,
        Tong Zhang,
        Jun Liu,
        Tingting~Yang,~\IEEEmembership{Member,~IEEE},
        Shanshan Song,
        and~Jun-Hong~Cui,~\IEEEmembership{Member,~IEEE}
\IEEEcompsocitemizethanks{
\IEEEcompsocthanksitem Yu Gou, Tong Zhang, Shanshan Song, and Jun-Hong Cui are with the College of Computer Science and Technology, Jilin University, Changchun, China, 130012. Jun-Hong Cui is also with Shenzhen Institute for Advanced Study, UESTC, Shenzhen, China. E-mail: \{gouyu18@mails, tongzhang18@mails, songss@,  junhong\_cui@\}.jlu.edu.cn.
\IEEEcompsocthanksitem Jun Liu is with the School of Electronic and Information Engineering, Beihang University, Beijing, China, 100191. E-mail: liujun2019@buaa.edu.cn.
\IEEEcompsocthanksitem Tingting Yang is with Department of Network Intelligence, Peng Cheng Laboratory, Shenzhen, China, and also with Navigation College, Dalian Maritime University, Dalian, China. E-mail: yangtingting820523@163.com.}
\thanks{(Corresponding author: Jun Liu.)}
}

%
%

\markboth{Journal of \LaTeX\ Class Files,~Vol.~14, No.~8, August~2015}%
{Shell \MakeLowercase{\textit{et al.}}: Bare Demo of IEEEtran.cls for Computer Society Journals}
%



\IEEEtitleabstractindextext{%
\begin{abstract}

This paper investigates the fair-effective communication and robustness in imperfect and energy-constrained underwater acoustic sensor networks (IC-UASNs). Specifically, we investigate the impact of unexpected node malfunctions on the network performance under the time-varying acoustic channels. Each node is expected to satisfy Quality of Service (QoS) requirements. However, achieving individual QoS requirements may interfere with other concurrent communications. Underwater nodes rely excessively on the rationality of other underwater nodes when guided by fully cooperative approaches, making it difficult to seek a trade-off between individual QoS and global fair-effective communications under imperfect conditions. Therefore, this paper presents a \underline{SE}mi-\underline{CO}operative \underline{P}ower \underline{A}llocation approach (SECOPA) that achieves fair-effective communication and robustness in IC-UASNs. The approach is distributed multi-agent reinforcement learning (MARL)-based, and the objectives are twofold. On the one hand, each intelligent node individually decides the transmission power to simultaneously optimize individual and global performance. On the other hand, advanced training algorithms are developed to provide imperfect environments for training robust models that can adapt to the time-varying acoustic channels and handle unexpected node failures in the network. Numerical results are presented to validate our proposed approach.
\end{abstract}

}

\maketitle

\IEEEdisplaynontitleabstractindextext

%
\IEEEpeerreviewmaketitle

\IEEEraisesectionheading{\section{Introduction}\label{sec:introduction}}

\IEEEPARstart{U}{nderwater} Acoustic Sensor Networks (UASNs) have been envisioned as a necessity to enable the space–air-ground–aqua integrated networks (SAGAIN) \cite{liu2020task}. To facilitate a broader range of underwater applications, UASNs are expected to provide fair, effective, and robust services. In addition to these visions, limited energy supplies and harsh underwater environments are two major concerns for UASNs. First, the underwater nodes are battery-powered and severely energy-constrained; it is incredibly challenging and expensive to replenish/replace the batteries after deployments \cite{mohammadi2020increasing}. Second, UASNs are unreliable due to the unique features of the acoustic channel, including high loss ratio and bit error rate, long and variable propagation delay, non-stationary channel conditions, and limited bandwidth \cite{stojanovic2009underwater}. Even worse, UASNs are imperfect due to unexpected node malfunctions and surrounding interference.

Due to the aforementioned challenges, the effectiveness of the network is far below expectations, and the role of UASNs in SAGAIN will be diminished. Since the underwater nodes are sparsely deployed and the propagation delay of acoustic signals is much larger than the packet transmission time \cite{noh2014dots}, researchers have taken advantage of spatial separation and have proposed multiple transmission strategies to increase the chances of concurrent transmissions while reducing the likelihood of collisions \cite{gupta2000capacity}. Some of these solutions avoid collisions between underwater nodes sharing the same acoustic channel by carefully scheduling transmissions or tuning transmission probabilities \cite{syed2007understanding}\cite{guan2012stochastic}\cite{li2015dtmac}\cite{zhong2020new}\cite{guan2020stochastic}. In these solutions, the transmission powers of the underwater nodes are generally assumed to be identical, and the successful reception of a transmission is modeled by the protocol model \cite{shi2012bridging}.

While in other types of solutions commonly used in UASNs, the power management-based methods, researchers enable multiple concurrent communications in the network through rational allocation of transmit power among nodes, thus cooperatively improving the network performance \cite{su2015joint}\cite{qian2016maca}\cite{zhen2016transmission}\cite{mohsan2020investigating}\cite{zhang2021udarmf}\cite{gou2021achieving}\cite{gou2022deep}. Power control is recognized as a practical means to optimize the overall system performance across the physical and the medium access control layers \cite{jornet2008distributed}. By adjusting power levels, underwater nodes can collaboratively satisfy their QoS requirements while avoiding destructive interference with other simultaneous communications. It is worth noting that these methods typically incorporate an implicit transmission scheduling mechanism by setting the transmit power of one or more underwater nodes to zero.

Fairness is always considered as an additional performance indicator in network performance optimization algorithms due to the potential consequences of unfair resource allocation, such as flow starvation \cite{sathiaseelan2007multimedia}. It is expected that the underwater nodes will have fair access to the shared acoustic channel, that their QoS requirements will be properly met, and that the link performance (e.g., energy consumption, link quality, and throughput) will be reasonable and fair \cite{huaizhou2013fairness}. As mentioned in \cite{diamant2016leveraging}, the aim of fairness is to guarantee equal transmission opportunities for nodes. When maximizing the performance of UASNs, the lack of fairness guarantee mechanisms will lead to heterogeneous transmission opportunities for different users \cite{luo2015efficient}. It has been demonstrated in \cite{xu2017fairness} that unfair transmission opportunities lead to heterogeneous delivery delays, thereby deteriorating the average network delivery delay as the variability of transmission opportunities between links increases.

Furthermore, UASNs are imperfect due to node malfunctions and acoustic entities in their vicinity (e.g., autonomous underwater vehicles and malicious interference adversaries). Over-reliance on cooperation mechanisms makes the system unstable and vulnerable \cite{zhao2022coach}. In addition, the level of interference depends on both concurrent communication and surrounding acoustic entities \cite{young2011overcoming}. Models that consider only intra-network interference may not be adaptable to real-world systems.

This paper presents SECOPA, a semi-cooperative power allocation approach to achieve fair-effective communication and robustness in IC-UASNs. The approach is based on distributed multi-agent reinforcement learning (MARL). To the best of our knowledge, we are the first to investigate communication fairness in imperfect and energy-constrained UASNs. We formulate the UASNs as multi-agent systems (MAS) and the underwater nodes as intelligent agents. The distributed MARL-based approach allows the agent to adapt its strategies to unreliable acoustic channels and imperfect networks. The main contributions of this paper are listed as follows.

\begin{enumerate}
\item We formulate the fair-effective communication and robustness optimization problems in IC-UASNs. We aim to (a) seek a better trade-off between individual QoS requirements and global fair-effective communications through power allocation in imperfect environments, and (b) increase the delivery ratio to avoid wasting energy. 

\item We define the network utility and then transform the optimization problem into maximizing long-term network utility. The power allocation task in IC-UASNs is conceptualized as a decentralized partially observed Markov decision process (Dec-POMDP), which is solved using a deep MARL-based algorithm and a delicately designed reward function. The devised reward function maximizes the long-term communication fairness and effectiveness, while reducing the number of fail transmissions to avoid energy waste. Furthermore, the power allocation model is trained through a centralized training with decentralized execution (CTDE) paradigm. Agents decide their transmission strategies based solely on local observations without exchanging information with other agents. Consequently, the cooperative mechanism of the proposed power allocation model does not introduce additional communication overhead to the system.

\item We propose SECOPA, a semi-cooperative approach to manage the power allocation tasks for achieving fair-effective communication and robustness in IC-UASNs. SECOPA consists of three components: (a) an imperfect-environment generator (IE-Generator), which comprises three advanced learning strategies and provides imperfect environments for training robust power allocation models; (b) a deep MARL-based power allocation model, which guides the underwater nodes to take individual actions that simultaneously achieve local QoS requirements and global fair-effective communications; and (c) a performance evaluator, which evaluates the model and instructs the IE-Generator to respond to model performance.

\item We conducted extensive simulations on IC-UASNs power allocation tasks to show the need to consider the unexpected failures in underwater nodes and the interference from surrounding acoustic entities. Numerical results have demonstrated the superiority and generalization of our proposed SECOPA over baseline methods in optimizing network performance, including network capacity, delivery ratio, communication fairness, and network reuse. Furthermore, the effects of two critical hyper-parameters (utility threshold and advanced learning factor) on the model performance have been investigated.

\end{enumerate}

The remainder of the paper is organized as follows. Section \ref{sec:rw} reviews the related work. Section \ref{sec:formu} gives the system model and the problem formulation. Section \ref{sec:method} describes the semi-cooperative power allocation approach (SECOPA). Section \ref{sec:exp} gives the evaluation results. Finally, Section \ref{sec:con} concludes the paper.

\section{Related Work}\label{sec:rw}


Recently, increasing the chances of simultaneous transmissions while reducing the probability of collisions in underwater communications has attracted considerable interest from both the academic and industrial communities. Some of these solutions avoid collisions between underwater nodes sharing the same acoustic channel by carefully scheduling transmissions or tuning transmission probabilities, while others enable multiple concurrent communications in the network by rationally allocating the transmit power of the nodes, thus cooperatively improving the network performance.

Solutions that fall into the first category often assume a homogeneous network in which the transmit power of the underwater nodes is identical and the successful reception of a transmission is characterized by the protocol model \cite{shi2012bridging}. First, in \cite{syed2007understanding}, the authors showed that larger guard bands in slotted networks can help reduce the probability of collisions. Later, in \cite{guan2012stochastic}, a queue-aware distributed access technique was presented to deal with spatial and temporal uncertainty. Each transmitter locally optimizes a transmission probability profile based on past observations to decide its transmission behavior. Based on the traditional protocol interference model, \cite{li2015dtmac} proposed a delay-tolerant transmission strategy that modifies the transmission probability to maximize throughput. \cite{zhong2020new} and \cite{guan2020stochastic} are two contemporary modeling techniques for acoustic channels. The former is based on the protocol model, while the latter is based on the physical model \cite{gupta2000capacity}. These two solutions improve the sum throughput of the network by allowing transmitters to independently modify their transmission probability profiles over a series of time slots.

Transmit power management strategies, on the other hand, facilitate more concurrent communications by adjusting the transmit power of transmitters to fulfill their local QoS requirements without interfering with other concurrent communications. In \cite{shi2012bridging}, the authors confirmed that the use of power control at each transmitting node can effectively improve the validity of protocol model-based methods. In \cite{su2015joint}, Su \textit{et al.} developed a dynamic transmit power adjustment algorithm to maximize the spatial reuse efficiency in UASNs and reduce the energy consumption. \cite{qian2016maca} emphasized that UASNs are energy-constrained systems, and it is necessary to consider the trade-off between node energy consumption and achievable performance. \cite{zhen2016transmission} and \cite{mohsan2020investigating} aimed to reduce energy consumption when optimizing network performance. The former seeks a trade-off between energy consumption and communication quality by allocating more power to communication links with strong signal strength, while allocating less power to links suffering from severe fading. The latter focuses on extending the lifetime of energy-constrained UWSNs while maintaining network throughput. Such methods typically implicitly incorporate a transmission scheduling mechanism by adjusting the transmit power of one or more underwater nodes to zero \cite{elbatt2004joint}.

In the aforementioned studies, the properties of acoustic channels, the inadequate spatial reuse of UASNs, and the fact that nodes are energy constrained have been thoroughly investigated; however, communication fairness has not been studied in depth. Fairness is an important and interdisciplinary concept that is used in numerous fields \cite{huaizhou2013fairness}. Fair allocation of transmission opportunities is especially important in distributed systems where a set of resources is to be shared by a number of users \cite{jain2008art}. UASNs are typical distributed and multi-user systems. The research community has devoted considerable attention to addressing the issue of maintaining communication fairness while simultaneously increasing the number of concurrent communications in UASNs. Xie and Cui developed an underwater transmission scheme that resolves packet collision and supports fairness in an energy-efficient manner \cite{xie2007r}. However, their solution supports fairness at the expense of channel utilization. In \cite{syed2008t}, the authors highlighted the space-time uncertainty and high latency of acoustic communication and proposed a distributed and energy-efficient transmission scheduling scheme. Taking into account the unique characteristics of acoustic channels, the proposed approach simultaneously optimizes channel utilization, energy consumption, and communication fairness. The above schemes either assume that all nodes transmit at the same power level or do not consider transmission power. Bouabdallah \textit{et al.} allocate different transmit power to underwater nodes, taking into account both the peculiarities of the acoustic channel and the limited energy supplies \cite{bouabdallah2017joint}. However, they aimed at overcoming the energy hole problem in UASNs and focused less on network reuse.

Due to their effective and efficient decision-making capabilities, reinforcement learning methods have become increasingly popular in underwater resource allocation tasks. Wang \textit{et al.} implemented a distributed resource allocation strategy for UASNs using cooperative Q-learning to optimize the total data rate while satisfying the communication quality requirements \cite{wang2019self}. Ye \textit{et al.} proposed a DRL-based transmission strategy for UWANs to maximize the network throughput by carefully utilizing the available time slots caused by long propagation delays or not used by other nodes \cite{ye2020deep}. To maximize local capacity and global concurrency, a distributed and adaptive resource management paradigm is presented in \cite{zhang2021udarmf}. However, UDARMF does not consider node mobility and communication fairness. \cite{gou2021achieving} aims to achieve time-sharing and spatial-reuse UWSNs with communication fairness, but the delivery ratio is not maximized. \cite{gou2022deep} developed a deep MARL-based power management strategy (DMPM) for increasing the fair reuse of UWSNs. While node mobility is considered in DMPM, the delivery ratio and the imperfect nature of the UASNs need further investigation.

The above RL-based solutions assume that UASNs are fully cooperative multi-agent systems. Sensor networks are basically cooperative multi-agent systems \cite{zhang2011scaling}. Due to the uncertainty in the environment and the partial observability of sensors, decision problems in sensor networks can be modeled as a Decentralized Partially Observable Markov Decision Process (DEC-POMDP). Cooperation among MARL agents enables them to outperform traditional RL agents when operating as a team, but also increases the vulnerability of a team to the failure of one of its constituent agents \cite{lin2020robustness}. Existing research on increasing the number of concurrent communications and improving communication fairness often assumes that underwater nodes can operate normally at all times, ignoring the imperfect nature of UASNs. In this paper, we present a semi-cooperative power allocation approach (SECOPA) that achieves fair-effective communication and robustness in IC-UASNs. To the best of our knowledge, we are the first to study communication fairness in IC-UASNs.

\section{System Model and Problem Formulation}\label{sec:formu}

\subsection{System Overview}\label{sec:system}

\begin{figure}[htbp]
\begin{center}
\includegraphics[width=3in]{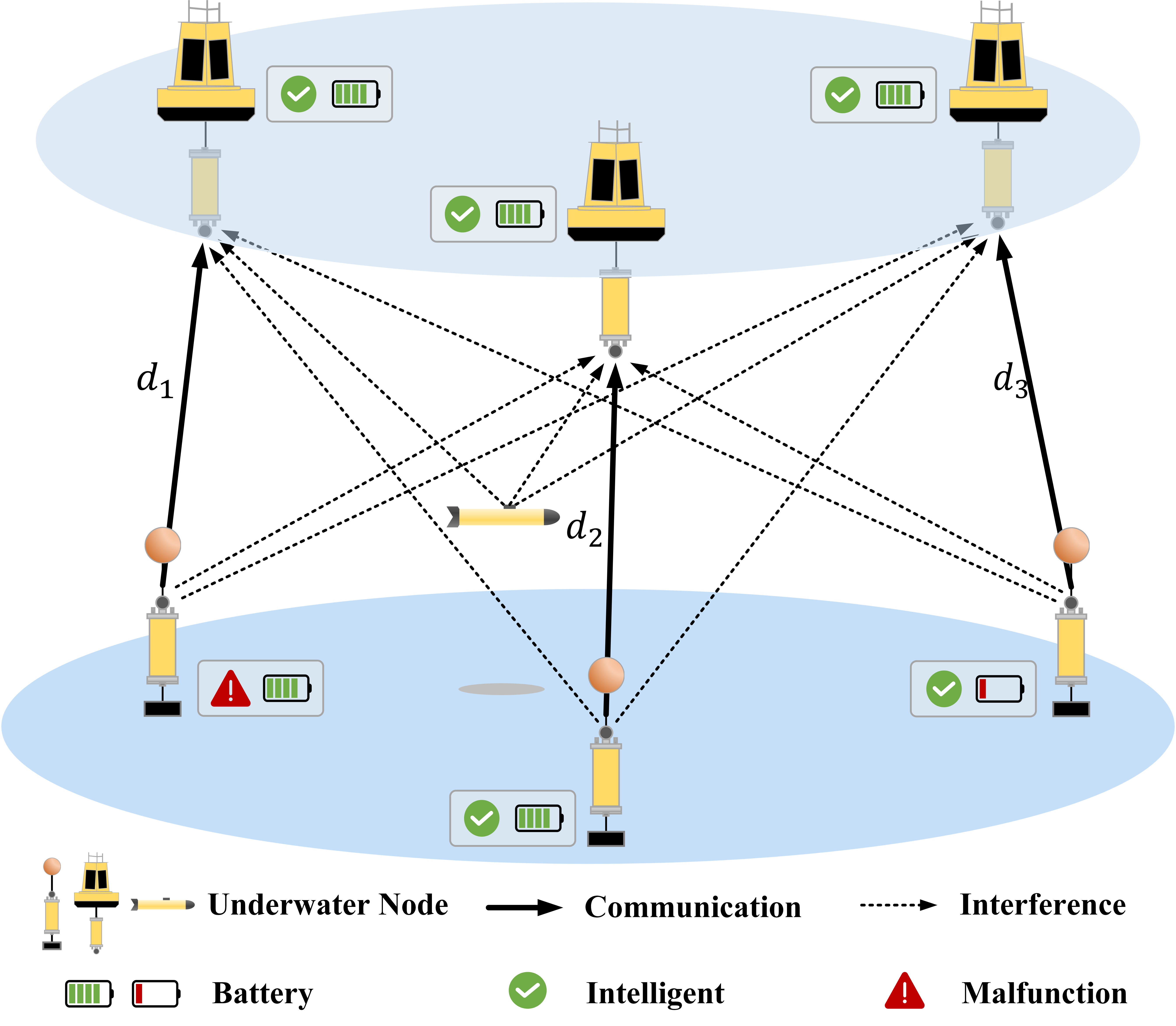}
\vspace{-1em}
\caption{An imperfect and energy-constrained UASNs (IC-UASNs) consists of multiple transmitter-receiver pairs. The $d_{i}$ ($i$=1,2,3) is the distance from $n_i$ to its intended receiver.}
\label{fig:imUWSN}
\end{center}
\end{figure}

\begin{figure*}[htbp]
\centerline{\includegraphics[width=6.5in]{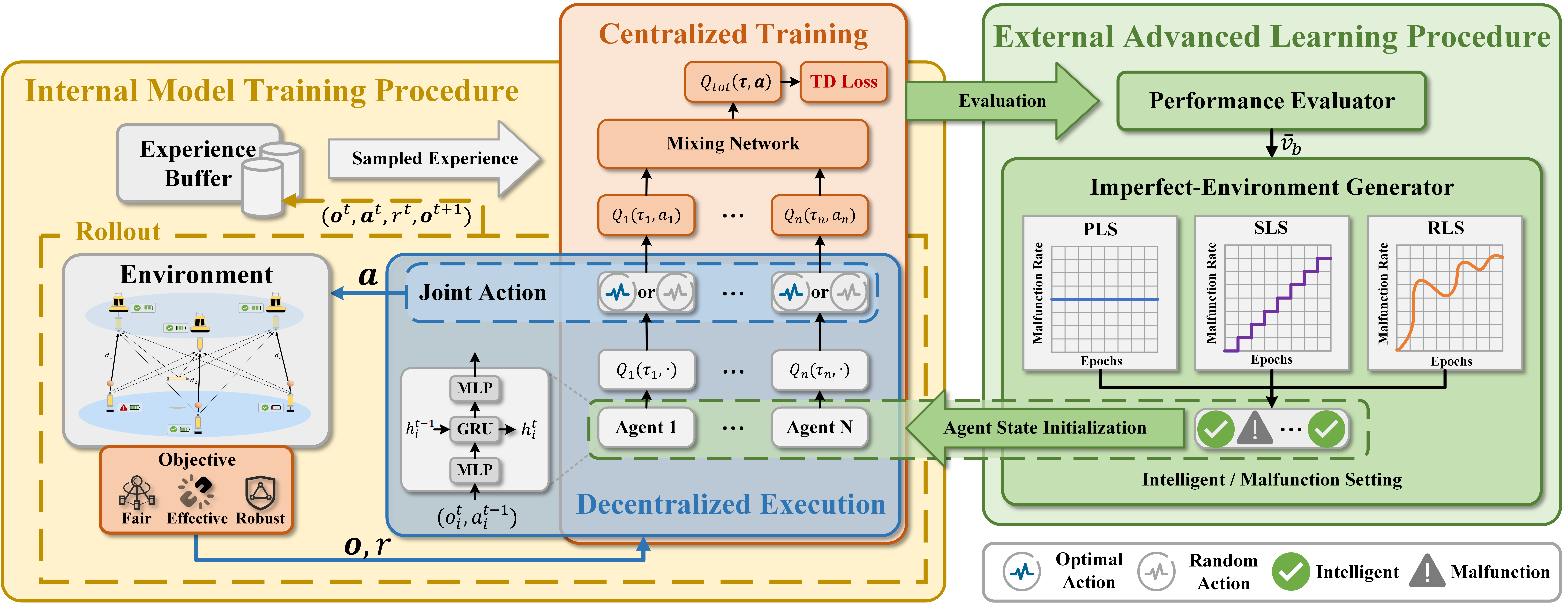}}
\vspace{-0.6em}
\caption{Flowchart of SECOPA. In the \textit{external advanced learning procedure}, the IE-Generator provides imperfect environments for model training and evaluation, and the model is evaluated by the performance evaluator. In the \textit{internal model training procedure}, the agents interact with the environment to train the power allocation model.}
\label{fig:main}
\end{figure*}

This paper considers a single-hop underwater acoustic sensor network with unexpected node malfunctions and constrained energy supplies to support various underwater applications, as shown in Fig. \ref{fig:imUWSN}. Let $\mathcal{N}\triangleq\{n|n=1,2,\ldots,N\}$ be the set of transmitters sharing the acoustic channel and equipped with half-duplex and omnidirectional acoustic modems. Assuming that the clock synchronization of all underwater nodes has been pre-configured prior to deployment using an external timing source, such as a laptop connected to the Internet that establishes communication with NTP servers to acquire precise time information for clock synchronization, and the nodes can maintain synchronization for a period of time \cite{chefrour2021one}. $N$=$|\mathcal{N}|$ is the number of transmitters. The transmitters send to their intended receiver in a time-slotted manner, in which time domain of the communication channel is divided into time slots and the transmitters send at the beginning each available slot \cite{song2019optimizing}\cite{ye2020deep}. Let $\mathcal{N}_{int} \subseteq \mathcal{N}$ be the subset of intelligent nodes, and $\mathcal{N}_{mal}\subseteq \mathcal{N}$ be the subset of nodes with unexpected malfunctions. $\mathcal{N}_{int} \cup \mathcal{N}_{mal}$=$\mathcal{N}$, $\mathcal{N}_{int} \cap \mathcal{N}_{mal}$=$\varnothing$. The intelligent transmitters coordinate to enable more concurrent transmissions and satisfy the QoS requirements, while the malfunctioning nodes behave irrationally (take random actions or stop transmission). The underwater nodes move passively due to ocean currents, tides and other factors, that have a moving speed of $c$ in meters per second (m/s), and the mobility model $\mathcal{M}_{k,c}$ is as given in \cite{he2020trust}. $\boldsymbol{x}_{i}^{t+1}$=$\mathcal{M}_{k,c}(\boldsymbol{x}_{i}^{t})$, where $\boldsymbol{x}_{i}^{t}$ denotes $n_i$'s coordinates at $t$ slot. Each transmitter $n_i$ is associated with a specific receiver $n_{\tilde{i}}$, and the communication link is denoted by $l_{i,\tilde{i}}$ if the distance $d_{i,\tilde{i}}$ between $n_i$ and $n_{\tilde{i}}$ is no more than $tr_{i}$ as (\ref{equ:link}), otherwise there is no communication link.
\begin{equation}
d_{i,\tilde{i}}=|\boldsymbol{x}_{i}-\boldsymbol{x}_{\tilde{i}}| \le tr_{i}
\label{equ:link}
\end{equation}
where $tr_{i}$ is the transmission range of $n_i$. Note that the approach proposed in this paper is not limited to any particular kinematic model. However, the model should at least consider the effect of water currents on underwater nodes. We then introduce three definitions regarding the communication links as follows:
\begin{defn}
Active link. $l_{i,\tilde{i}}$ is active when $n_i$ sends to $n_{\tilde{i}}$ at time slot $t$.
\label{defn:activelink}
\end{defn}
\begin{defn}
Effective link. $l_{i,\tilde{i}}$ is effective when $n_i$'s transmission is successfully received at $n_{\tilde{i}}$.
\label{defn:effectivelink}
\end{defn}
\begin{defn}
Active-but-ineffective link. $l_{i,\tilde{i}}$ is active-but-ineffective when $n_i$ sends to $n_{\tilde{i}}$ at time slot $t$, but the transmission is not successfully received at $n_{\tilde{i}}$.
\label{defn:ineffectivelink}
\end{defn}

An effective link is also an active one, while the active one may not be effective due to interference. Table \ref{tab:para} summarizes the notations. For readability, we sometimes drop notation for time slot $t$.
\begin{table}[htp]
\caption{Notation Description.}
\begin{center}
\vspace{-1em}
\begin{tabular}{p{0.73cm}<{\centering}lp{0.73cm}<{\centering}l}
\hline
Notation&Description&Notation&Description\\
\hline

$\mathcal{N}$&Transmitter set&$N$&Number of transmitters\\
$n_i$,$\tilde{n_i}$&Transmitter, receiver&$l_{i,\tilde{i}}$&Communication link\\
$s_{i,\tilde{i}}$&Behavior indicator&$re_{i,\tilde{i}}$&Behavior indicator\\
$p$&Transmit power&$\mathcal{P}$&Power set\\
$\gamma$&Received SINR&$\gamma^{0}$&$\gamma$ threshold\\
$\delta^{0}$&Required lifetime&$\delta_{dur}$&Slot duration\\
$f_c$&Carrier frequency&$B$&Bandwidth\\
$c_{i,\tilde{i}}$&Data rate&$e_i$/$E_i$&Energy consumption\\
$\mathcal{C}_{\mathcal{N}}$&Network capacity&$\varphi_{\mathcal{N}}$&Fairness index\\
$\mathcal{R}_{\mathcal{N}}$&Network reuse&$\mathcal{W}_{\mathcal{N}}$&Failed communications\\
$\mathcal{U}_{\mathcal{N}}$&Network utility&$\kappa_{i}$&Malfunction state\\
\hline
\end{tabular}
\end{center}
\label{tab:para}
\end{table}
\vspace{-1em}

\subsection{Physical Channel Model}\label{sec:model}

The physical channel model accounts for transmission loss, multi-path fading, ambient noise, and interference from simultaneous transmissions and surrounding acoustic entities. The Signal-to-Interference-plus-Noise Ratio (SINR) at the receiver is defined as the power of the intended signal divided by the sum of the interference power (from all the other interfering signals) and the power of some ambient noise, denoted as $\gamma_{i,\tilde{i}}$ in this paper and calculated as (\ref{equ:SINR}) \cite{stojanovic2007relationship}: 
\begin{equation}
\gamma_{i,\tilde{i}}=\frac{\eta_{0} p_{i}g_{i,\tilde{i}}(d_{i,\tilde{i}},k,f_c)}{\eta_{0}\sum_{j \in \mathcal{N} \backslash i} p_{j}g_{j,\tilde{i}}(d_{j,\tilde{i}},k,f_c)+I_{s}+I_{a}}
\label{equ:SINR}
\end{equation}
where $\eta_{0}$ is the transducer efficiency in converting electrical to acoustic power (which generally ranges from 80 to 90 percent \cite{wills2006low}), $p_{i},p_{j} \in \mathcal{P}$ are the transmit power of $n_{i}$ and $n_j$. $\mathcal{P}$ is the set of all available transmit power for the underwater nodes, $P\!=\!|\mathcal{P}|$ is the number of available transmit power. $g_{i,\tilde{i}}$ and $g_{j,\tilde{i}}$ are the channel gains of $l_{i,\tilde{i}}$ and $l_{j,\tilde{i}}$, which are related to the time-varying distance $d_{i,\tilde{i}}$ and $d_{j,\tilde{i}}$, spreading factor $k$, and carrier frequency $f_c$. For $l_{i,\tilde{i}}$, the channel gain is calculated as (\ref{equ:gain}),
\begin{equation}
g_{i,\tilde{i}}=G_{i,\tilde{i}}\rho^{2}
\label{equ:gain}
\end{equation}
where $\rho$ is the fading coefficient, which is a function of time and can be approximated by using a unit-mean Rayleigh distributed random variable with a cumulative distribution function expressed as $P[\rho\le x]=1-\exp(\frac{\pi x^{2}}{4})$ \cite{guan2020stochastic}. $G_{i,\tilde{i}}$ is the transmission loss experienced by a narrow-band-acoustic signal over a given spectrum, and can be described by the Urick propagation model as (\ref{equ:gain2}) \cite{urick1975principles},
\begin{equation}
G_{i,\tilde{i}}=(1000\times d_{i,\tilde{i}})^{-k}\cdot10^{-\frac{\alpha(f_c)d_{i,\tilde{i}}+A_{mn}}{10}}
\label{equ:gain2}
\end{equation}
where $k$ describes the geometry of propagation, $k=1$ for cylindrical spreading, $k=1.5$ for practical spreading (adopted in this paper), and $k=2$ for spherical spreading; $d_{i,\tilde{i}}$ in km is the distance between the $n_{i}$ and its intended receiver; $A_{mn}$ is the transmission anomaly in decibels, which accounts for the loss of acoustic intensity due to multi-path propagation, refraction, diffraction, and scattering of acoustic signals \cite{pompili2009cdma}; $\alpha(f_c)$ is the absorption coefficient as (\ref{equ:Thorp}), which gives $\alpha(f_c)$ in dB/km and $f_c$ in kHz \cite{stojanovic2007relationship}.
\begin{equation}
\begin{split}
10\log \alpha(f_c)=&0.11\frac{f_c^{2}}{1+f_c^{2}}+44\frac{f_c^{2}}{4100+f_c^{2}}\\&+2.75\cdot10^{-4}f_c^{2}+0.003
\label{equ:Thorp}
\end{split}
\end{equation}
For lower frequencies that above a few hundred Hz, the absorption coefficient is calculated as (\ref{equ:lowgain})
\begin{equation}
10\log \alpha(f_c)=0.11\frac{f_c^{2}}{1+f_c^{2}}+0.011f_c^{2}+0.002
\label{equ:lowgain}
\end{equation}
$\eta_{0}\sum_{j \in \mathcal{N} \backslash i} p_{j}g_{j,\tilde{i}}(d_{j,\tilde{i}},k,f_c)$ represents the cumulative interference strength experienced by the intended receiver caused by other transmitters. $I_{s}$ denotes the interference from surrounding acoustic entities, such as Autonomous Underwater Vehicles (AUVs). The identification of interference signals is similar to \cite{bai2015link}. Furthermore, $I_{a}=N(f_c)\bigtriangleup f$ in (\ref{equ:SINR}) represents the noise power of the underwater acoustic channel. $N(f_c)$ is the total power spectral density (PSD) in dB re $\mu$Pa per Hz of the ambient noise, which is composed of turbulent noise, transportation noise, thermal noise, and wave noise, as expressed in \cite{stojanovic2007relationship}.$\bigtriangleup f$ is a narrow band around frequency $f_c$. This paper employs the Rayleigh fading model due to its effectiveness in both shallow and deep water scenarios and its simplicity \cite{yang2009optimization}, while other fading models (e.g., Rician, TWDP, and Nakagami-$m$) may also be applicable to the considered system.


\subsection{Slot Model}\label{sec:slotModel}

In slotted-UASNs, time is divided into slots, and the transmitters send at the beginning of each allocated slot \cite{guerra2009performance}. Due to the slow propagation speed and severe multi-path propagation, underwater communications suffer from significant propagation delays and multi-path delay spread (i.e., the difference in delays between the first and last significant multi-path arrivals). To avoid inter-slot interference, the slot duration must consider the propagation delay and delay spread in the guard interval \cite{morozs2020channel}. Similar to\cite{morozs2020channel} and \cite{zhu2014toward}, the slot duration $\delta_{dur}$ is defined as (\ref{equ:slot}),
\begin{equation}
\delta_{dur}=\delta_{tran}+\max_{i \in \mathcal{N}}\{\delta^{prop}_{i,\tilde{i}}+\delta^{ds}_{i,\tilde{i}}\}
\label{equ:slot}
\end{equation}
where $\delta_{tran}$ is the transmission duration, $\delta^{prop}_{i,\tilde{i}}$ and $\delta^{ds}_{i,\tilde{i}}$ refer to the propagation delay and multi-path delay spread between $n_i$ and its intended receiver, respectively.

In simulation-based research on UASNs, two widely employed approaches for determining the propagation delay $\delta^{prop}_{i,\tilde{i}}$ are the Euclidean distance-based method and the BELLHOP-based method. The former calculates the propagation delay based on the Euclidean distance between the transmitter-receiver pair and a fixed propagation speed \cite{stojanovic2007relationship}. The latter simulates the realistic impulse responses of multiple acoustic channels using BELLHOP \cite{porter2011bellhop} beam tracing, generating a set of parameters (such as propagation delay and delay spread) of the received signals with the simulated channel realizations \cite{morozs2020channel}. Since the underwater environment is time-varying and acoustic signals may not travel in straight lines between the transmitter-receiver pairs, this paper adopts the BELLHOP-based method to simulate the acoustic propagation delay $\delta^{prop}_{i,\tilde{i}}$ and delay spread $\delta^{ds}_{i,\tilde{i}}$. $\delta_{tran}$ is determined based on the application design. The slot duration is determined offline during deployment or modified online through control signals.


\subsection{UASNs Constraints}\label{sec:cons}

\textbf{Energy constraints.} Most underwater nodes are battery-powered and thus have limited energy supplies. Let $e_{i}(t)=p_{i}^{t}\times \delta_{tran}$ in joule (J) be the energy consumed by $n_i$ at time slot $t$, associated with an executed transmit power $p_{i}(t)$. $E_{i}(\delta^{t})=\sum_{t=1}^{\delta^{t}}e_{i}(t)$ represents the total energy consumption of $n_i$ during the past $\delta^{t}$ time slots. For underwater applications, a certain network lifetime $\delta^{0}$ is required to accomplish the tasks. In this paper, the \textit{network lifetime} is defined as the duration until the first intelligent node becomes inoperable due to battery depletion \cite{mohammadi2020increasing}. Since the network lifetime primarily relies on the initial battery capacity, denoted as $E^{0}$, as well as the energy consumption of the underwater nodes, it follows that $\forall i \in \mathcal{N}_{int}$, the energy consumption over the required network lifetime should not exceed the maximum available battery capacity, i.e., $E_{i}(\delta^{0})=\sum_{t=1}^{\delta^{0}}e_{i}(t) \delta_{tran}) \le E^{0}$.

\textbf{Communication threshold.} In this paper, \textit{Physical Model} \cite{gupta2000capacity} is used to model the successful reception of a transmission over $l_{i,\tilde{i}}$, where a minimum $\gamma$ is required for successful reception. When the received SINR $\gamma_{i,\tilde{i}}$ exceeds a threshold $\gamma^{0}$, the transmission from $n_{i}$ can be decoded by its intended receiver $n_{\tilde{i}}$ with an acceptable bit error rate \cite{zhong2020new}, i.e., $\gamma_{i,\tilde{i}} \ge \gamma^{0}$. Substituting (\ref{equ:SINR}) and (\ref{equ:gain}) into the above inequality, a successful transmission from $n_{i}$ to $n_{\tilde{i}}$ can be expressed as:
\begin{equation}
\gamma_{i,\tilde{i}}=\frac{\eta_{0} p_{i}G_{i,\tilde{i}}\rho^{2}}{\eta_{0}\sum_{j \in \mathcal{N} \backslash i} p_{j}G_{j,\tilde{i}}\rho^{2}+I_{s}+I_{a}} \ge \gamma^{0}
\label{equ:detailgamma}
\end{equation}
Assuming slowly time-varying and quasi-static channels \cite{jing2017energy}, the distance between any two nodes remains unchanged at each time slot, and is allowed to change in the subsequent slot according to the mobility model. From (\ref{equ:detailgamma}), the desired transmit power $p_{i}$ should satisfy (\ref{equ:detailgamma2}):
\begin{equation}
p_{i} \ge \gamma^{0} \frac{\eta_{0}\sum_{j \in \mathcal{N} \backslash i} p_{j}G_{j,\tilde{i}}\rho^{2}+I_{s}+I_{a}}{\eta_{0}G_{i,\tilde{i}}\rho^{2}}
\label{equ:detailgamma2}
\end{equation}
The joint transmit power scheme $P_{\mathcal{N}}$=$[p_1,\dots,p_{N}]$ for a system with $N$ simultaneous communications should satisfy the $N$ inequalities as (\ref{equ:detailgamma2}). As the number of nodes increases, the strong levels of interference (generated by simultaneous communications) cannot be overcome by power control\cite{elbatt2004joint}, which is sometimes destructive and cannot be ignored for acoustic communications. In order to facilitate more effective communications and enhance the delivery ratio, it is essential to schedule the active links and allocate appropriate transmit power for these links at each time slot. For successful communications, the transmit power for the scheduled active links should satisfy the above inequality, i.e., the communication threshold is met in the presence of both ambient noise and communication interference.

Let $c_{i,\tilde{i}}(t)$ be the achievable data rate in bits-per-second (bps) of $l_{i,\tilde{i}}$ at time slot $t$, and $B$ is the bandwidth. When $\gamma_{i,\tilde{i}}(t) < \gamma^{0}$, $c_{i,\tilde{i}}(t)=0$; when $\gamma_{i,\tilde{i}}(t) \ge \gamma^{0}$, $c_{i,\tilde{i}}(t)$ is calculated as Shannon formula, i.e., $c_{i,\tilde{i}}(t)=B\times \log_{2}(1+{\gamma}_{i,\tilde{i}}(t))$.

\begin{defn}
Network capacity $C_{\mathcal{N}}$ is defined as the total amount of data transmitted over all links during the network lifetime, and can be calculated as (\ref{equ:netCapacity}).
\begin{equation}
C_{\mathcal{N}}=\sum_{i=1}^{N}\sum_{t=1}^{\delta}(c_{i,\tilde{i}}(t) \times \delta_{tran})
\label{equ:netCapacity}
\end{equation}
\label{defn:capacity}
\end{defn}
\vspace{-1.5em}

Let $s_{i,\tilde{i}}(t)$ be the binary indicator denotes whether $n_i$ transmit at time slot $t$. $s_{i,\tilde{i}}(t)=1$ when $l_{i,\tilde{i}}$ is active at time slot $t$, and $s_{i,\tilde{i}}(t)=0$ otherwise. Let $re_{i,\tilde{i}}(t)$ indicates whether $n_{\tilde{i}}$ received from $n_i$ at time slot $t$. $re_{i,\tilde{i}}(t)=1$ when $l_{i,\tilde{i}}$ is effective, $re_{i,\tilde{i}}(t)=0$ otherwise.

\begin{defn}
Network reuse $\mathcal{R}_{\mathcal{N}}$ is defined as the number of effective communications at each transmission slot as $\mathcal{R}_{\mathcal{N}}(t)=\sum_{i=1}^{N}re_{i,\tilde{i}}(t)$, and $\mathcal{R}_{\mathcal{N}}^{t}$ is the average reuse during the past $t$ slots, as (\ref{equ:avereuse}).
\begin{equation}
\mathcal{R}_{\mathcal{N}}^{t}=\frac{1}{t}\frac{1}{N}\sum_{t=1}^{t}\mathcal{R}(t)=\frac{1}{t}\frac{1}{N}\sum_{t=1}^{t}\sum_{i=1}^{N}re_{i,\tilde{i}}(t)
\label{equ:avereuse}
\end{equation}

\label{defn:reuse}
\end{defn}
\vspace{-1.8em}

\subsection{Fairness and Effectiveness}

The underwater networks are multi-user systems, and all nodes should be treated fairly. In this paper, we use Jain's fairness index for evaluation \cite{jain2008art}, which defines fairness as a function of the variety of deliveries among underwater nodes. Communication fairness needs to be considered in almost all resource allocation problems in wireless networks, without which resource starvation or energy holes may occur \cite{huaizhou2013fairness}. To calculate the average communication fairness, the \textit{sliding window method} is utilized \cite{berger2004short}. Given different horizons of the window, communication fairness $\varphi_{\mathcal{N}}$ is categorized as follows:

\begin{defn}
Lifetime-horizon Fairness (FR-LH). FR-LH evaluates the fairness along the past $t$ slots or the entire network lifetime, the window size increasing with $t$. 
\label{defn:lf}
\end{defn}

\begin{defn}
Adaptive-horizon Fairness (FR-AH). FR-AH evaluates the system fairness over past $\alpha$ time slots. $\alpha$ is determined by network deployment.
\label{defn:af}
\end{defn}

The fairness is calculated as (\ref{equ:Fair1}) and (\ref{equ:Fair2}).
\begin{equation}
\varphi_{\mathcal{N},h}(t)=\frac{(\sum_{i=1}^{N}\sum_{t-h+1}^{t}re_{i,\tilde{i}}(t))^{2}}{N\sum_{i=1}^{N}(\sum_{t-h+1}^{t}re_{i,\tilde{i}}(t))^{2}}, h \ge t
\label{equ:Fair1}
\end{equation}
\begin{equation}
\varphi_{\mathcal{N},h}(t)=\frac{(\sum_{i=1}^{N}\sum_{t=1}^{t}re_{i,\tilde{i}}(t))^{2}}{N\sum_{i=1}^{N}(\sum_{t=1}^{t}re_{i,\tilde{i}}(t))^{2}}, 0<t<h
\label{equ:Fair2}
\end{equation}
where $h$ is the horizon size. $h$=$t$ for FR-LH and $h$=$\alpha$ for FR-AH. $\varphi_{\mathcal{N}} \in [0,1]$. At each time slot, if $k$ nodes make successful transmissions where the other $(N\!-\!k)$ nodes fail, $\varphi_{\mathcal{N}}$=$k/N$. Specifically, if all nodes communicate equally and successfully with their intended receivers, $\varphi_{\mathcal{N}}$=$1$; if a single node transmits through the acoustic channel, the network is extremely unfair, $\varphi_{\mathcal{N}}$=$1/N$; if no node can make successful communications, the network fails, $\varphi_{\mathcal{N}}=0$.

Effective communication is a prerequisite for achieving network fairness. When $s_{i,\tilde{i}}(t)\!=\!1$ and $re_{i,\tilde{i}}(t)\!=\!1$, the current transmission is successful without wasting energy, which increases the network capacity and is the \textit{best-case}. When $s_{i,\tilde{i}}(t)\!=\!1$ and $re_{i,\tilde{i}}(t)\!=\!0$, all the energy used for this transmission is wasted while causing interference to other nodes in the network, which is the \textit{worst-case}. In between is $s_{i,\tilde{i}}(t)\!=\!0$ and $re_{i,\tilde{i}}(t)\!=\!0$, where $n_i$ does not participate in the transmission at this time slot, and does not contribute to or harm the network performance. We should increase concurrent communications and avoid wasting limited energy supplies. $\mathcal{W}_{\mathcal{N}}(t)$ counts the average number of active-but-ineffective links during the past $t$ slots, as given in $\mathcal{W}_{\mathcal{N}}^{t}\!=\!\frac{1}{t}\frac{1}{N}\sum_{t=1}^{\delta}\sum_{i=1}^{N}|s_{i,\tilde{i}}(t)-re_{i,\tilde{i}}(t)|$.

\subsection{Node Malfunctions}

In underwater systems, the nodes that pre-configured with identical intelligent algorithms to cooperatively improve network performance may experience malfunctions due to hardware or software failures. The hardware failures render the underwater nodes completely unavailable, while the software failures may lead to irrational transmission behaviors. This paper focuses primarily on software-related malfunctions, specifically the cases where the cooperative optimization algorithms are not loaded. Each node, denoted as $n_i \!\in\! \mathcal{N}$, presents a probability $\epsilon$ of failing to load the intelligent cooperative optimization algorithm. To capture the state of malfunction for node $n_i$, a binary variable $\kappa_{i}\!\in\! \{0,1\}$ is introduced. Specifically, $\kappa_{i}\!=\!1$ indicates that $n_i$ is experiencing a malfunction, while $\kappa_{i}\!=\!0$ signifies its normal operation. In this paper, we assume that $\kappa_{i}$ follows a Bernoulli distribution with a parameter of $\epsilon$. With $\epsilon$, the intelligent node degrades into a malfunction node that behaves irrationally (i.e., takes random actions or stops transmitting).

Furthermore, the intelligent algorithm relies on observation information as input, which may either be unavailable or missing during certain time intervals. This unavailability or absence of observation data can lead to the failure of the intelligent algorithm, thereby weakening the cooperation mechanism among the nodes and diminishing the overall network utility. Note that the malfunction node cannot recover its functionality, whereas the nodes unable to execute intelligent algorithms due to missing observations can regain their availability if the observations are received correctly. As the probability $\epsilon$ increases, the number of intelligent nodes losing their functionalities (which behave irrationally instead of following the pre-configured cooperative optimization algorithm) also increases. In such cases, the difficulty of providing fair-effective communication increases accordingly.

\subsection{Problem Formulation}\label{sec:probfor}

The maximization of long-term network utility $\mathcal{U}_{\mathcal{N}}$ is of critical importance when constrained energy supplies and spatial-temporal fluctuations in channel conditions are considered. In this paper, $\mathcal{U}_{\mathcal{N}}$ includes fairness utility and reuse utility, which is defined as,
\begin{equation}
\mathcal{U}_{\mathcal{N}}=\varphi_{\mathcal{N},h}(t)+\mathcal{R}_{\mathcal{N}}^{t}
\label{equ:utility}
\end{equation}
where the fairness utility $\varphi_{\mathcal{N},h}(t) \!\in\! [0,1]$, reuse utility $\mathcal{R}_{\mathcal{N}} \!\in\! [0,1]$, and thus $\mathcal{U}_{\mathcal{N}} \!\in\! [0,2]$. With the system model and underwater constraints presented in the above sections, the optimization problem can be formulated as (\ref{equ:objmax})-(\ref{equ:conPower}),
\begin{subequations}
\begin{align}
\max_{\mathbf{P}}\quad& \varphi_{\mathcal{N},\delta}(\delta) +\mathcal{R}_{\mathcal{N}}^{\delta} \label{equ:objmax}\\
\mbox{s.t.}\quad
& \gamma_{i,\tilde{i}}(t) \ge \gamma^{0}, \forall i\in \mathcal{N}_{int}, \forall t\in [0,\delta]\label{equ:conSINR}\\
& \delta_{i} \ge \delta^{0}, \forall i\in \mathcal{N}_{int}\label{equ:conLife}\\
& E_{i}(t) \le E^{0}, \forall i\in \mathcal{N}, \forall t\in [0,\delta]\label{equ:conEnergy}\\
& re_{i,\tilde{i}}(t) \in \{0,1\}, \forall i\in \mathcal{N}, \forall t\in [0,\delta]\label{equ:conReuse}\\
& \sum_{i=1}^{N}\kappa_{i} \in[0,N], \forall i\in \mathcal{N}\label{equ:conMal}\\
& p_{i}(t) \in \mathcal{P}, \forall i\in \mathcal{N}, \forall t\in [0,\delta]\label{equ:conPower}
\end{align}
\end{subequations}
where $\mathbf{P}\!=\![\boldsymbol{p}_{i}^{\top}]_{i=1}^{N} \!\in\! \mathbb{R}^{N \!\times\! \delta}$ is the matrix of transmit powers performed by the underwater nodes at each time slot, and $\boldsymbol{p}_{i}$ is the transmit power performed by $n_i$ at each time slot during the network lifetime $\delta$. Constraint (\ref{equ:conSINR}) is the communication quality constraint to ensure that the communication from intelligent node is effective. Constraint (\ref{equ:conLife}) provides guarantee for satisfied network lifetime. Constraint (\ref{equ:conEnergy}) bounds that the total consumed energy should not exceed the battery capacity at each node. Nodes that deplete energy supplies are not allowed and unable to transmit. Constraint (\ref{equ:conReuse}) presents the restriction for the acoustic modem that no more than one transmission can be successfully received at each receiver at each transmission slot. Constraint (\ref{equ:conMal}) limits the number of malfunction nodes. In the best case, all intelligent nodes maintain their functionality. In the worst case, all intelligent nodes experience software-related failures and degrade into malfunction nodes that behave irrationally. In other cases, the intelligent nodes and malfunction nodes coexist in the system. Constraint (\ref{equ:conPower}) ensures that the performed transmit power is available for the adopted acoustic modem. The optimal transmit power allocation scheme $\boldsymbol{p}^{*}$ is decided as (\ref{equ:MAXutility}).
\begin{equation}
\boldsymbol{p}^{*}=\arg\max_{{\mathbf{P}}}\mathcal{U}_{\mathcal{N}}
\label{equ:MAXutility}
\end{equation}

Since we are evaluating the fairness of concurrent communications, we consider the cases where the transmitters are (1) \textit{greedy}, meaning that they always have data to transmit, and (2) \textit{homogeneous}, meaning that all transmitters have identical software/hardware configurations and benefit from similar channel conditions. No transmitter is disadvantaged by its transmit power, battery capacity, or deployment; shadow zone and convergence zone are outside the scope of our investigation.

\section{Proposed Solution}\label{sec:method}

To address the problem described in $\S$\ref{sec:probfor}, we present SECOPA, a semi-cooperative reinforcement learning-based approach that achieves fair-effective communication and robustness in IC-UASNs through power allocation. In order to strike a balance between individual QoS and global network utility, the intelligent nodes first decide whether to transmit at each time slot, and then select the optimal transmit power based on local observations. Through the implicit joint link scheduling and power allocation scheme, SECOPA achieves the optimization objective. As shown in Fig. \ref{fig:main}, SECOPA consists of three components: imperfect-environment generator (IE-Generator), power allocation model, and the performance evaluator. In $\S$\ref{sec:ats}, we offer three advanced learning strategies to implement the IE-Generator, which simulates the unexpected malfunctions and surrounding acoustic interference for training robust models. $\S$\ref{sec:drlFrame} describes the DRL-based power allocation model, and $\S$\ref{sec:detailFrame} gives the details.

\subsection{Advanced Learning Strategies for Unexpected Malfunctions}\label{sec:ats}

SECOPA’s IE-Generator is responsible for providing imperfect environments (with unexpected malfunctions and surrounding acoustic interference) to train a robust power allocation model. We offer three advanced learning strategies for implementing the IE-Generator, i.e., prior knowledge-based learning strategy (PLS), syllabus-based learning strategy (SLS), and responsive learning strategy (RLS). The power allocation models, which will be introduced later, are trained in such kinds of environments to resolve unanticipated node malfunctions. According to different learning strategies, the IE-Generator builds the training environment with a malfunction rate of $\epsilon$, $\epsilon \in [0,\epsilon^{u}]$. By parallel with human learning approaches, this research seeks to identify training strategies for robust power allocation models. Following are the three advanced training strategies.
\begin{enumerate}[a)]
\item \textbf{Prior knowledge-based learning strategy (PLS)} provides environments with a predetermined malfunction rate $\epsilon$ based on prior knowledge from field experimentation. PLS anticipates that agents will be able to maintain network robustness under typical imperfect conditions. The PLS procedures are provided in Algorithm \ref{alg:PLS}.

\begin{algorithm}
\caption{Procedures of Prior Knowledge-based Learning Strategy (PLS)}\label{alg:PLS}
Initialize $episode \gets 1$\\
Initialize malfunction rate $\epsilon$, update cycle $n_{uc}$, model evaluation times $n_{me}$, evaluation network $\theta$, episode reward $R$, mean episode reward $\bar{R}$\\
\For{$episode \gets 1,2,\ldots,M$}
{ 
\If{$episode$ / $n_{uc}$==0}{
/* Start the $\theta$ evaluation process */\\
Train the model according to Algorithm \ref{alg:modelTraining} and update $\theta \gets \theta_{episode}$.\\
With $\theta$, evaluate the model for $n_{me}$ times\\
Calculate $\bar{R} \gets \frac{1}{n_{me}}\sum_{i=1}^{n_{me}}R(\theta,i)$ \\
}
\Else{
Train the model according to Algorithm \ref{alg:modelTraining} and update $\theta \gets \theta_{episode}$.\\
}
Finish current episode\\
Update $episode \gets episode+1$
}
\Return $\theta \gets \arg\max\bar{R}(\theta_{episode},n_{me})$

\end{algorithm}

\begin{algorithm}
\caption{Procedures of Syllabus-based Learning Strategy (SLS)}\label{alg:SLS}
Initialize malfunction rate $\epsilon \gets 0$, $episode \gets 1$\\
Initialize upper bound $\epsilon^{u}$, update cycle $n_{uc}$, evaluation network $\theta$,  step size $\triangle \epsilon = \frac{\epsilon^{u} \cdot n_{uc}}{M}$\\
\For{$episode \gets 1,2,\ldots,M$}
{
\If{$episode$ / $n_{uc}$==0}{
/* Start the $\epsilon$ update process */\\
Increase $\epsilon$ by $\epsilon \gets \epsilon+\triangle \epsilon$\\
}
\Else{
Keep $\epsilon$ constant\\
}
Train the model according to Algorithm \ref{alg:modelTraining} and update $\theta \gets \theta_{\epsilon}$.\\
Finish current episode\\
Update $episode \gets episode+1$\\
}
\Return $\theta$

\end{algorithm}

\begin{algorithm}
\caption{Procedures of Responsive Learning Strategy (RLS)}\label{alg:RLS}
Initialize malfunction rate $\epsilon \gets 0$, $episode \gets 1$,  utility performance $v_{b} \gets 0$, \\
Initialize utility threshold $v_{b}^{0}$, upper bound $\epsilon^{u}$, advanced learning factor $\Gamma$, evaluation network $\theta$, update cycle $n_{uc}$, model evaluation times $n_{me}$, episode reward $R$, mean episode reward $\bar{R}$\\
\For{$episode \gets 1,2,\ldots,M$}
{	
\If{$episode$ / $n_{uc}$==0}{
/* Start the $\epsilon$ update process */\\
\If{$\bar{v_{b}} \ge v_{b}^{0}$}{
\If{$\epsilon > \epsilon^{u}$}{
Bound $\epsilon$ by $\epsilon=\epsilon^{u}$
}
\Else{
Increase $\epsilon$ by $\epsilon=\epsilon+\Gamma(1-\epsilon)$
}
}
\Else{
\If{$\epsilon+\Gamma(0-\epsilon)<0$}{
Bound $\epsilon$ by $\epsilon=0$
}
\Else{
Decrease $\epsilon$ by $\epsilon=\epsilon+\Gamma(0-\epsilon)$
}
}
Train the model according to Algorithm \ref{alg:modelTraining} and update $\theta \gets \theta_{episode}$.\\
With $\epsilon$ and $\theta$, evaluate the model for $n_{me}$ times\\
Calculate $\bar{v_{b}} \gets \frac{1}{n_{me}}\sum_{i=1}^{n_{me}}v_{b}(\theta,\epsilon,i)$ and $\bar{R} \gets \frac{1}{n_{me}}\sum_{i=1}^{n_{me}}R(\theta,i)$\\
}
\Else{
Train the model according to Algorithm \ref{alg:modelTraining} and update $\theta \gets \theta_{episode}$.\\
}
Finish current episode\\
Update $episode \gets episode+1$\\
}
\Return $\theta \gets \arg\max\bar{R}(\theta_{episode},n_{me})$

\end{algorithm}

\item \textbf{Syllabus-based learning strategy (SLS)} aims to scale the malfunction rate with the number of learning epochs. SLS trains the model by incessantly increasing $\epsilon$, and thus has greater generalizability in real deployments. The SLS procedures are described in Algorithm \ref{alg:SLS}.

\item \textbf{Responsive learning strategy (RLS)} modifies $\epsilon$ in response to agent performance. RLS generates an environment with an increased $\epsilon$ when the agents' performance in the current environment exceeds a predetermined threshold. In contrast, RLS decreases $\epsilon$ when the agents do not meet the utility threshold. Procedures of RLS are presented in Algorithm \ref{alg:RLS}. 

\end{enumerate}
We preset the utility threshold $v_{b}^{0}$ for Algorithm \ref{alg:RLS} to evaluate and guide node behaviors. With a higher utility threshold, the node is expected to perform better in current situation. A learned or adaptive threshold is left for future development.

\subsection{DRL-based Power Allocation Framework}\label{sec:drlFrame}

The joint optimization problem of fair-effective communication and robustness in IC-UASNs can be modeled as a decentralized partially observed Markov decision process (Dec-POMDP) consisting of a tuple $G\!=\!\langle \mathcal{S}, \mathcal{A}, P, \mathcal{R}, O, \mathcal{O}, n, \gamma \rangle$ \cite{bernstein2002complexity}. $s \!\in\! \mathcal{S}$ is the environment state, $\mathcal{A}$ is the action set. Since the agent is partially observable, it can only make local observation $o_{i} \in \mathcal{O}$, following the observation function $O(s,i)$. At each time slot, the agent selects an action $a_{i} \!\in\! \mathcal{A}$ according to policy $\pi$, $\boldsymbol{a}$ is the joint action. The environment transits to the next state according to the transition function $P(s^{t+1}|s^{t},\boldsymbol{a})$ and the agents are rewarded by $r\!=\!\mathcal{R}(s,\boldsymbol{a})$.

\textbf{Problem Transformation.} Deep MARL is a model-free method for solving Dec-POMDP. Before solving the formulated problem with a deep MARL algorithm, we must at first construct an underwater communication-specific action space $\mathcal{A}$, observation set $\mathcal{O}$, and reward function $\mathcal{R}$. As mentioned in (\ref{equ:MAXutility}), we seek to identify the optimal policy $\pi^{*}$ that increases the number of fair-effective communications and optimizes network robustness.

\textbf{Action Space $\mathcal{A}$.} $\mathcal{A}$ is constructed by the node transmit power set $\mathcal{P}$. Since the nodes considered in this paper are identical, $\mathcal{A}=\mathcal{A}_{i}, i \in \mathcal{N}$. For an intelligent node, the executed action is the one that leads to the largest reward based on its partial observations, $a_{i}(t)=\arg\max Q(\boldsymbol{o}_{i}(t), \boldsymbol{a})$. The malfunction nodes take random actions. $\boldsymbol{a}(t)=[\boldsymbol{a_{i}}(t)]_{i=1}^{N} \in \mathbb{R}^{N}$ is the joint action at time slot $t$.

\textbf{Observation Space $\mathcal{O}$.} $\mathcal{O}$ is the set of partial observations made by the agent to make decisions. At slot $t$, $\boldsymbol{o}_{i}(t)\!=\!\boldsymbol{o}_{i}^{L}(t)\!\cup\!\boldsymbol{o}_{i}^{I}(t)\!\cup\! \boldsymbol{o}_{i}^{D}(t)$ is $n_i$'s observations. $\boldsymbol{o}_{i}^{L}(t)\!=\!\{d_{i}(t),e_{i}^{re}(t),c_{i,\tilde{i}}(t),r_{i,\tilde{i}}^{to}(t),a_{i}(t-1),i\}$ consists of $n_{i}$'s local information, including position $d_{i}(t)$, residual energy indicator $e_{i}^{re}(t)$, last transmission indicator $re_{i,\tilde{i}}(t)$, transmission ratio $r_{i,\tilde{i}}^{to}(t)$, last transmit power $a_{i}(t\!-\!1)$, and its identifier $i$. $\boldsymbol{o}_{i}^{I}(t)\!=\!\boldsymbol{o}_{i}^{Ic}(t)\!\cup\! \boldsymbol{o}_{i}^{Is}(t)\!=\!\{\boldsymbol{d}_{j}(t), \boldsymbol{d}_{s}(t), I_{s}\}$ is the observations about other transmitters and surrounding acoustic entities. $\boldsymbol{o}_{i}^{D}(t)\!=\!\{\boldsymbol{d}_{\tilde{i}}(t)\}$ is the about the intended receiver.

\textbf{Reward Function $\mathcal{R}$.} The reward function evaluates the executed actions and provides guidance for future decisions. In this paper, a team reward $r(t)$ is given to all agents at each time slot. We design two distinct types of reward functions. The first is similar to the network utility function $\mathcal{U}_{\mathcal{N}}$ as defined in $\S$\ref{sec:probfor}, and the action that optimizes both lifetime-horizon fairness and effective communications will be rewarded more, denoted as FR-LH and defined in (\ref{equ:FRLH}).
\begin{equation}
r(t)=\varphi_{\mathcal{N},L}(t) \times \sum_{i=1}^{N}re_{i,\tilde{i}}(t)
\label{equ:FRLH}
\end{equation}
In the second class, we penalize the fail transmissions with a penalty term $\varsigma$ for effective communication, $\varsigma$ is calculated as (\ref{equ:varsigma}). 
\begin{equation}
\varsigma=\frac{1}{N}\times \sum_{i=1}^{N}(s_{i,\tilde{i}}(t)-re_{i,\tilde{i}}(t))
\label{equ:varsigma}
\end{equation}
And to analyze how different horizons impact the learning process, we at first implement the reward functions with FR-LH and FR-AH as Definition \ref{defn:lf} and \ref{defn:af}, and then calculate the average number of effective communications throughout their horizons. These reward functions are denoted as E-FR-LH and E-FR-AH, calculated as (\ref{equ:EFRLH}) and (\ref{equ:EFRAH}), respectively.
\begin{equation}
r(t)=\varphi_{\mathcal{N},L}(t) \times \sum_{i=1}^{N}re_{i,\tilde{i}}(t)-\varsigma
\label{equ:EFRLH}
\end{equation}

\begin{equation}
r(t)=\varphi_{\mathcal{N},A}(t) \times \sum_{i=1}^{N}re_{i,\tilde{i}}(t)-\varsigma
\label{equ:EFRAH}
\end{equation}
Specifically, if the QoS requirements are not met, each agent receives a -100 penalty and the episode ends. The optimization objective is to maximize the expected cumulative discounted reward $\mathbb{E}[\sum_{t=1}^{\delta}\gamma r(t)]$. Let $\gamma \!\in\! [0,1]$ be the discount factor that adjusts the agent's attention to future rewards.

\subsection{Centralized Training and Decentralized Execution of the SECOPA}\label{sec:detailFrame}

\begin{algorithm}
\caption{Training algorithm for SECOPA}\label{alg:modelTraining}
Initialize experience buffer $\mathcal{B}$ to capacity $B_r$;\\
Initialize each agent's evaluation network $\theta$ and target network $\theta^{-}$, $\theta^{-} \gets \theta$, C.\\

With probability $\epsilon$, $\kappa_{i} \gets 1$, otherwise $\kappa_{i} \gets 0$.\\
\For{$\delta \gets 1, ..., \delta^{0}$}{
\For{$i \gets 1, ..., N$}{

\If {$E_{i}(t)\le E^{0}$} 
{
\If {$\kappa_{i}==0$}{
With probability $\varepsilon$ randomly select $a_{i}(t) \in \mathcal{A}$\\
otherwise $a_{i}(t) \gets \arg\max_{\boldsymbol{a_{i}}} Q_i(\tau_{i}(t),\boldsymbol{a_{i}};\theta)$.\\
}
\Else{
Select $a_{i}(t)\in \mathcal{A}$ following guard policy $\pi_{d}$ or $\pi_{s}$\\
}
}
\Else{
$\delta^{0}$ is not satisfied, $\delta \gets t$\\
Finish current episode.\\}
}
Execute joint action $\boldsymbol{a}(t)$, receive reward $r(t)$ and new observation $\boldsymbol{o}(t+1)$ of all agents.\\
Create and add transition to $\mathcal{B}$.\\
Update $t \gets t+1$\\
}
Randomly select $b$ transitions from $\mathcal{B}$, train the evaluation network $\theta$ with the mini-batch data.\\
\For{$j \gets 1,2,...,b$}
{
Set $y_{j} \gets r_{j}$ if episode terminates at $j+1$ slot\\
otherwise set $y_{j}$ as (\ref{equ:up}).\\
Update $\theta$ by minimizing $Loss(\theta)$ as (\ref{equ:td}) and (\ref{equ:upTheta}).\\
}
\If {episode / C == 0}{
$\theta^{-} \gets \theta$
}
Finish current episode.\\
\Return $\theta_{episode}$
\end{algorithm}

SECOPA utilizes the centralized training with decentralized execution (CTDE) paradigm to effectively learn policies for agents. All agents perceive and act in the environment so as to maximize their shared utility. Due to the limitations of real-world data collection, i.e., sampling inefficiency and cost, simulation environments are used to train the various agents \cite{zhao2020sim}. In the CTDE paradigm, training is performed using a simulator with additional state information, but the agents must rely on local action-observation histories during execution \cite{foerster2016learning}. The training data is generated and stored in a finite-size experience buffer from the continuous interaction between the agents and the simulator following a particular behavior policy (e.g., $\varepsilon$-greedy policy). Value-decomposition network (VDN), which learns a decentralized utility function for each agent and uses a mixing network to combine these local utilities into a global action value, is one of the promising ways to exploit the CTDE paradigm \cite{sunehag2018value}. We abstract a centralized virtual controller that combines the actions of each agent and distributes the global reward to them. For centralized training, the central controller learns only one network (i.e., the power allocation model), which is then used by all agents in distributed UASNs environments. However, agents with identically learned models may still behave differently because they receive different observations and thus evolve distinct hidden states. During the decentralized execution of the learned policies, the central controller is removed, and each agent uses its own copy of the learned network, evolving its own hidden state, and selecting its own actions based on its own local observations \cite{foerster2016learning}. The SECOPA workflow is outlined below:
\begin{enumerate}[a)]
\item Set learning strategies for centralized training process.
\item Train the model and update network parameters.
\item Execute the underwater power allocation tasks with properly trained model.
\end{enumerate}

As shown in Fig. \ref{fig:main}, the centralized training phase contains two interactive procedures: 1) external advanced learning procedure, and 2) internal model training procedure. In the external procedure, the IE-Generator provides imperfect environments for model training and evaluation, and the model is evaluated by the performance evaluator. Strategies for implementing the IE-Generator have been presented in $\S$\ref{sec:ats}. In the internal procedure, the agents interact with the provided environment to train the power allocation model. The virtual controller mixes the Q-value of all agents and calculates a global reward for the agents. Through identical feedback signals, agents are aware of how the environment reacts to their executed actions, as opposed to how much their actions contribute to the joint reward. Since the agent uses the deep Q-network (DQN) structure \cite{mnih2015human}, the updates for the evaluation and target networks are asynchronous. The internal procedure updates the evaluation network, whereas the external procedure updates the target network.

In the centralized training phase, we seek to identify the optimal deterministic policy $\pi$ that directs the agent to obtain the largest expected cumulative reward through individual operations at each time slot. Since the state space for underwater networks is unlimited, the conventional tabular-based solution is inadequate for underwater power allocation tasks. In this paper, we parameterize the value function $Q_{\theta}(s,a)$ to approximate the action-value in discrete action space, i.e., $Q_{\theta}(s,a)\simeq Q_{\pi}(s,a)$, where $\theta$ are the parameters of the approximation function and are updated by reinforcement learning. We construct a DQN-based agent network to address instabilities induced by the experience replay mechanism, which randomizes over the data to remove the correlations in the observations and smooth out changes in the data distribution. In addition, the target network is updated only periodically in the external procedure, reducing correlations with the target \cite{mnih2015human}.

Only agents that satisfy the energy constraints in (\ref{equ:conEnergy}) are allowed to transmit at each time slot. According to malfunction states, the agents are divided into two categories. The first are the intelligent agents, which cooperate under the guidance of SECOPA to maximize the global reward at each time slot. If the input information is temporarily unavailable to the intelligent nodes at certain time slots, the node will take random actions. Nodes of the second type are unable to load their model due to software-related failures; they behave irrationally (i.e., take random actions or stop transmission) during each time slot regardless of application requirements and global utility. In this paper, we focus primarily on situations where the malfunction nodes can still transmit under a random scheme. Future research will focus on cases where malfunction nodes completely stop transmitting.

Algorithm \ref{alg:modelTraining} gives the training procedure. We represent agent networks with Deep Recurrent Q-network (DRQN), which replaces the first post-convolutional fully-connected layer in DQN with a recurrent unit to save memory and deal with partial observability \cite{hausknecht2015deep}. The recurrent unit can be either a gated recurrent unit (GRU), or be a long short-term memory (LSTM) unit, which take current partial observations $o_{i}(t)$ and previous action $a_{i}(t-1)$ as inputs. The agent network parameterizes its individual action-value function with parameters $\theta$ to evaluate $Q_{i}(\boldsymbol{\tau_{i}(t)},a_{i}(t);\theta)$, and with parameters $\theta^{-}$ to predict $Q_{i}(\boldsymbol{\tau_{i}(t+1)},a_{i}(t+1);\theta^{-})$. Each agent randomly selects an action with a probability of $\varepsilon$, and selects the action that maximizes Q-value with a probability of $1-\varepsilon$. The goal of the $\varepsilon$-greedy policy is to reconcile the trade-off between exploitation and exploration, so that the agent can both strengthen the evaluation of the actions it already knows to be good and explore new actions.

Let $\boldsymbol{a}(t)=\{a_{i}(t)|i\in \mathcal{N}\}$ be the joint action of all agents at $t$ slot, and $\boldsymbol{\tau}(t)=\{\boldsymbol{\tau}_{i}(t)|i\in \mathcal{N}\}$ be the joint observations. The mixing network $f(\cdot)$ gets all MARL agents' outputs and makes a mixture of them, which is a joint action-value function approximator as (\ref{equ:mix}). 
\begin{equation}
Q_{total}(\boldsymbol{\tau},\boldsymbol{a})=f(Q_{1}(\tau_{1}, a_{1}),...,Q_{n}(\tau_{n}, a_{n}))
\label{equ:mix}
\end{equation}
SECOPA is not restricted to any particular mixing structure if it can reflect the cumulative effect of the joint actions on the environment. VDN decomposes the team value function into agent-wise value functions and satisfies the individual-global maximization (IGM) constraint as (\ref{equ:IGM}) by summing up individual value functions $Q_{i}(\tau_{i}, a_{i})$ for each agent $i \in \mathcal{N}$ as (\ref{equ:Qtotal}). VDN has the same number of model parameters as independent learning because the summation structure of the mixing network does not require any additional parameters.

\begin{equation}
\arg\max_{\boldsymbol{a}}Q_{total}(\boldsymbol{\tau},\boldsymbol{a})=
\begin{pmatrix}
\arg\max_{a_{1}}Q_{1}(\tau_{1}, a_{1})\\
...\\
\arg\max_{a_{n}}Q_{n}(\tau_{n}, a_{n})
\end{pmatrix}
\label{equ:IGM}
\end{equation} 
\begin{equation}
Q_{total}(\boldsymbol{\tau}, \boldsymbol{a})=\sum_{i}Q_{i}(\boldsymbol{\tau}_{i}, a_{i})
\label{equ:Qtotal}
\end{equation}

After performing a joint action $\boldsymbol{a}(t)$ of agents in simulator, the agents receive an identical team reward $r(t)$ and observe the subsequent environment state. For each time slot $t$, the experience buffer $\mathcal{B}$ stores the transition $e^{t}=\{\boldsymbol{o}(t), \boldsymbol{a}(t), r(t), \boldsymbol{o}(t+1)\}$, and has a capacity of $|\mathcal{B}|$. We apply Q-learning updates by sampling a mini-batch of transitions of size $b$ from $\mathcal{B}$ at the end of a training episode. During the $j$-th update, the MARL agents are trained in an end-to-end manner to minimize the time-difference loss as (\ref{equ:td}), and the model is updated with the set of parameters that minimizes $Loss(\theta)$ as (\ref{equ:upTheta}).
\begin{equation}
Loss(\theta)=\sum_{j}^{b}(y_{j}^{total}-Q_{total}(\boldsymbol{\tau^{t}}, \boldsymbol{a^t}; \theta))^2
\label{equ:td}
\end{equation}
\begin{equation}
y_{j}^{total}=r_{j}+\gamma max{Q(\boldsymbol{\tau^{t+1}}, \boldsymbol{a^{t+1}};\theta^{-})}
\label{equ:up}
\end{equation}
\begin{equation}
\theta \gets \arg\min_{\theta}Loss(\theta)
\label{equ:upTheta}
\end{equation}
where $\gamma$ is the discount factor that determines the look-back horizons, $\theta$ is the parameters of the evaluation network, $\theta^{-}$ is the parameters of the target network. $\theta^{-}$ is updated as (\ref{equ:upTheta}) every $C$ episodes, otherwise it stays unchanged between individual updates.

We encapsulate the optimal model (i.e., the one with the largest mean episode reward) for decentralized execution. The solid blue square in Fig. \ref{fig:main} shows the execution procedure. The IE-Generator, performance evaluator, and mixing network are removed. During the execution phase, intelligent underwater nodes independently decide whether to transmit at each time slot and their power $a_{i}(t)$ based exclusively on local partial observations $o_{i}(t)$, whereas malfunction nodes perform randomly at each time slot. The trained MARL agents decide their transmission strategies based solely on local observations without exchanging information with other agents. Therefore, the cooperative mechanism of the proposed power allocation model does not increase the system's communication overhead. Nevertheless, SECOPA relies on communication feedback, where the receiver informs the transmitter about the received signal strength of the previous transmission. The transmitter can adjust the transmission parameters accordingly. Consequently, the communication overhead predominantly arises from the transmission of the feedback, typically requiring an additional 4-6 bytes in real-world deployments \cite{di2019carma}.

Given an environment state, each agent calculates the Q-value for all available actions and selects the action with the highest Q-value. Assuming that the observation space has dimensions of $d_{o}$ and the action space has dimensions of $d_{a}$, the computational complexity \cite{ye2021scalable} for each agent is determined to be $O(d_{o}d_{a})$. In a MAS comprising $N$ agents, the total computational complexity scales linearly as $O(Nd_{o}d_{a})$. This linear rather than exponential relationship with the number of agents makes the proposed method implementable in real-world deployment scenarios. Note that it is imperative for the network topology to remain consistent between the training phase and the execution phase. Further research should be conducted to develop a performance optimization model that demonstrates enhanced generalization capabilities towards dynamic network topologies.

\subsection{Practicality of SECOPA}\label{sec:practicality}

Training DRL algorithms with data generated through simulations and using them to optimize underwater network performance in a distributed manner may raise concerns about the practicality issues caused by the sim-to-real gap. The method proposed in this paper reduces the sim-to-real gap in the following ways. \textit{First}, when constructing the simulator to comprehensively characterize the acoustic communication propagation process, SECOPA considers acoustic channel characteristics such as propagation delay, transmission loss, multi-path fading, ambient noise, and concurrent communication interference. The propagation delay and the realistic impulse responses of multiple acoustic channels are simulated through BELLHOP, which is a widely used channel modeling method in the underwater acoustic communications research community \cite{morozs2020channel}. By incorporating these factors into the simulator, we aim to mitigate the gap between simulation and real-world acoustic communication scenarios.

\textit{Second}, SECOPA considers the limited energy supply of the underwater nodes when optimizing the network performance, ensuring that the control policy obtained by the DRL algorithm is more practical and sustainable in real-world deployments. \textit{Third}, in practical underwater networks, one or more nodes may unexpectedly fail after deployment due to hardware or software failures. In recognition of this fact, SECOPA explicitly accounts for unexpected node failures in its optimization process. Instead of assuming that all nodes work perfectly at all times, our method strengthens the robustness of the model by accounting for possible node malfunctions after deployment. \textit{In addition}, SECOPA constrains the transmission behavior of nodes according to practical acoustic modem designs. Specifically, it considers discrete power levels available for transmission and aligns the node transmission behavior with the limitations and capabilities of real acoustic modems commonly used in underwater networks. By incorporating these practical constraints into our model, SECOPA effectively bridges the gap between simulation and reality. The effectiveness and robustness of SECOPA is validated in $\S$\ref{sec:exp}.

\section{Performance Evaluation}\label{sec:exp}

\subsection{Evaluation Setup}

\begin{figure*}
\begin{center}
\includegraphics[width=6in]{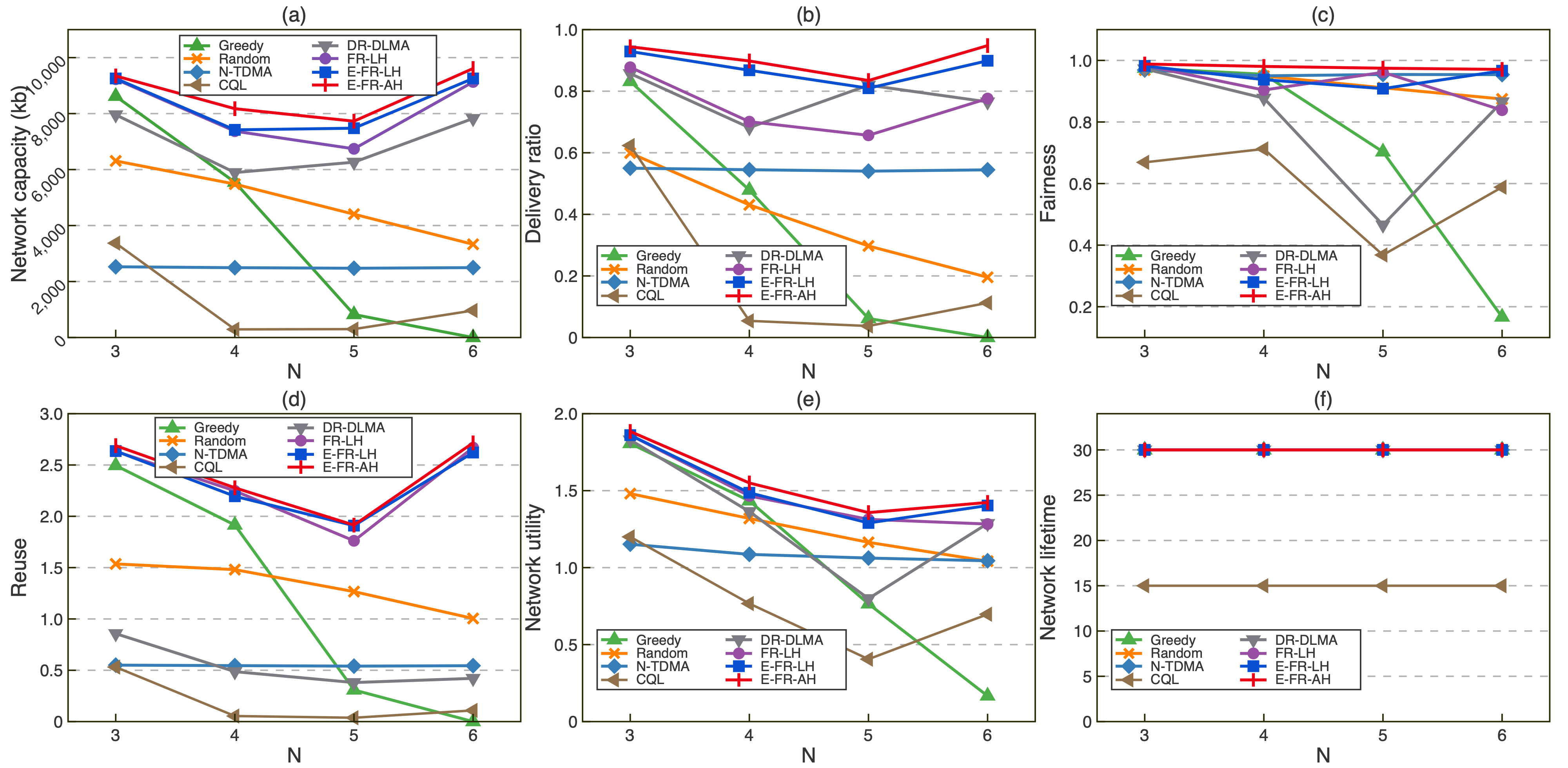}
\vspace{-1em}
\caption{Effects of SECOPA and baseline methods under different numbers of transmitter-receiver pairs on: (a) network capacity, (b) delivery ratio, (c) fairness, (d) reuse, (e) network utility, (f) network lifetime.}
\label{fig:noCrashCompare}
\end{center}
\end{figure*}
This section presents the numerical results of the performance of the proposed SECOPA. First, we give the simulation parameters, baseline methods, and evaluation metrics. Then, we present the performance of SECOPA in terms of throughput, communication fairness, network reuse, delivery ratio, delivery delay, and network utility of the network topology as shown in Fig. \ref{fig:imUWSN}.

This paper explores the fair-effective communication in the network. Therefore, we include multiple transmitter-receiver pairs with the same battery capacity and acoustic modem in the scenario. In addition, we deploy these transmitters and receivers evenly on the top surface and bottom of the cylindrical region to ensure that no transmitter is disadvantaged by deployments. The region has a radius of 1.4 km and a height of one kilometer. In real-world systems, the  acoustic communication systems are often interfered by surrounding mobile acoustic entities. In this paper, a mobile node is used to mimic spatio-temporal interference fluctuations to verify the adaptability of SECOPA in changing communication environments as \cite{wang2019self}. Localization, navigation, and path planning of the mobile node are beyond the scope of this work. The mobile node enters the region randomly and exits along a predetermined path, and the transmit power is set to 2 W according to \cite{wills2006low}. The parameters of the kinetic model to simulate the inevitable dynamic fluctuations of the nodes in underwater environment are set as \cite{he2020trust}, including the moving speed and the movement direction. $\delta_{tran}$ is set to 5 seconds, and the guard interval for propagation delay and delay spread of the acoustic channels are decided according to the BELLHOP beam tracing. In this paper, we aim to explore the saturation point \cite{noh2014dots} of the network capacity for a given topology and communication environment. The transmitters perform under a full buffer traffic model as \cite{morozs2020channel}. At the beginning of each time slot, transmitters can communicate with their designated receivers, while receivers receive but do not transmit data. Communication is effective when $\gamma^{0}$ is satisfied. When the remaining energy is less than 10 percent, the underwater nodes halt transmission. The adaptive evaluation horizon $\alpha$ for E-FR-AH is equal to $N$. The ambient noise power is calculated according to the model described in \cite{stojanovic2007relationship} with a 0 shipping activity level (low), wind speed $w\sim N(10,0.1)$ m/s. Table \ref{tab:paraset} lists the simulation parameters.
\begin{table}[htp]
\caption{Parameters.}
\vspace{-1em}
\begin{center}
\begin{tabular}{p{0.9cm}<{\centering}ll}
\hline
Parameter&Meaning&Value\\
\hline
$p$&Transmit power&[0, 2, 4, 8, 16, 32, 64] W\\
$E^{0}$&Battery capacity&5,000 J\\
$N$&Number of transmitters&[3,4,5,6]\\
$\eta_{0}$&Transducer efficiency&90\%\\
$\delta^{0}$&Network lifetime requirement&30 time slots\\
$f_{c}$&Carrier frequency&25 kHz\\
$B$&Bandwidth&5 kHz\\
$\gamma^{0}$&Communication threshold&10 dB\\
$\epsilon$&Malfunction rate&[0.01,0.1,0.2]\\
$v_{b}^{0}$&Utility threshold&1.25\\
$\Gamma$&Advanced learning factor&0.01\\
\hline
\end{tabular}
\end{center}
\label{tab:paraset}
\end{table}

The evaluation and target networks are constructed using two fully connected (FC) layers and a GRU. The first FC-layer consists of 64 hidden units, followed by a GRU with 64 hidden units and a ReLU to abstract the observations. The second FC-layer consists of seven hidden units that output Q-values for each possible action. The experience buffer capacity is 10,000, and the mini-batch size is 32. All networks use ADAM optimizer with a learning rate of 0.0005. The model was trained on 200,000 episodes. Every 200 episodes, the model was evaluated by executing 20 times with different initial random environments (i.e., different mobile node trajectories, node malfunction states, and node locations) and averaging the mean episode rewards. The optimal model is the one with the largest average reward. $\varepsilon$-greedy behavior policy is adopted to maintain a balance between exploration and expeditions for training, where $\varepsilon$ decreases linearly from 1 to 0.05 for the first 100,000 episodes and remains constant at $\varepsilon$=$0.05$ for the final 100,000 episodes. Such settings allow the agents to fully explore the environment in the early stages and gradually converge in the later stages to be able to cooperatively maximize long-term returns \cite{mnih2015human}. The discount factor $\gamma$ is set to 0.99. We compare SECOPA with five baselines.
\begin{enumerate}[i)]
\item \textbf{Greedy} \cite{yu2018power} guides all nodes to transmit simultaneously and uniformly at the maximum available power. All decisions are made based on energy constraints.
\item \textbf{Random} scheme instructs the nodes to select an arbitrary power at each time slot, regardless of the QoS requirements. Nodes are allowed to refrain from transmitting during some time slots.
\item \textbf{N-TDMA} scheme enables $N$ nodes to share the same bandwidth by dividing the acoustic channel into $N$ time slots. Nodes gain access to the channel in turn. Once a node completes its time slot, it must wait for the next allocated slot.
\item \textbf{CQL} \cite{wang2019self} aims to improve network capacity by scheduling communication links and allocating transmit power based on cooperative Q-learning method.
\item \textbf{DR-DLMA} \cite{ye2020deep} aims to maximize network capacity by reasonably utilizing the available time slots resulted from propagation delays or not used by other nodes.
\end{enumerate}
To verify the effectiveness of the proposed method, we evaluate the long-term performance by fairness $\varphi_{\mathcal{N}}$ as (\ref{equ:Fair1}) and (\ref{equ:Fair2}), network reuse $\mathcal{R}$ as (\ref{equ:avereuse}), delivery ratio $\mathcal{S}=\sum_{t=1}^{\delta}\sum_{i=1}^{N}\frac{re_{i,\tilde{i}}(t)}{s_{i,\tilde{i}}(t)}$, delivery delay $\mathcal{D}=\frac{(1-\mathcal{R})\times N \times \delta}{\mathcal{R}\times N \times \delta}=\frac{1-\mathcal{R}}{\mathcal{R}}$, and network utility $\mathcal{U}_{\mathcal{N}}$ as (\ref{equ:utility}).

\subsection{Results on Transmission Strategies}

Before evaluating the robustness of SECOPA in IC-UWSNs, we first validate the effectiveness of SECOPA's transmission strategy in UASNs with different network density and compare it with baseline methods. The number of transmitter-receiver pairs gradually increases from three to six. We implement SECOPA with FH-LH, E-FR-LH, and E-FR-AH, and give the comparison with five baseline methods in a UASN where $\epsilon$=0 and contains a mobile node. Figure \ref{fig:noCrashCompare} shows the effects of SECOPA and baseline methods under different numbers of transmitter-receiver pairs on network capacity, (b) delivery ratio, (c) fairness, (d) network reuse, (e) network utility, and (f) network lifetime.

The Greedy scheme is most affected by network density. When $N$=3, Greedy demonstrates superior performance across all performance metrics compared to Random and N-TDMA. It also achieves a level of performance that is comparable to the three implementations of SECOPA. As $N$ gradually increases, the performance of Greedy deteriorates rapidly. Specifically, when $N$ increases to 5, the performance of Greedy falls behind that of the Random and N-TDMA schemes. In the case of $N$=6, the network fails, wasting energy and communication resources used for transmission purposes. The performance of Random tends to degrade as the network density increases. And because the nodes do not consider whether the transmission was successful, the delivery ratio is inadequate. On the other hand, in N-TDMA, nodes access the channel alternately, with only one node transmitting during each time slot. Consequently, communication interference is caused solely by the surrounding mobile nodes. As a result, network density has no impact on network capacity, delivery ratio, communication fairness, and reuse of N-TDMA. As shown in Fig. \ref{fig:noCrashCompare}(d), the network reuse is less than one because the presence of mobile node renders certain time slots unusable.

The two learning-based baseline methods, i.e., CQL and DR-DLMA, perform differently in all scenarios. CQL assumes that each node operates independently to maximize local performance, without prioritizing cooperative efforts to enhance global utility. Consequently, as the network density increases, CQL faces increasing challenges in optimizing network performance. As $N$ increases to 5, CQL achieves similar performance as Greedy. Notably, as depicted in Fig. \ref{fig:noCrashCompare}(f), CQL consistently fails to meet the lifetime requirement. These findings underscore the necessity of a cooperative mechanism among underwater nodes to effectively optimize the overall network performance. In contrast, DR-DLMA attains performance comparable to SECOPA implementations. However, since DR-DLMA mainly focuses on maximizing network capacity without explicitly considering communication fairness and delivery ratio, its network utility is marginally inferior to that of SECOPA implementations. Additionally, as observed in Fig. \ref{fig:noCrashCompare}(a)-(e), DR-DLMA experiences more pronounced performance fluctuations as network density increases when compared to the three SECOPA implementations. This suggests that DR-DLMA may be more sensitive to network topology changes.

FR-LH, E-FR-LH, and E-FR-AH are SECOPA implementations with different reward functions. It is observed that models trained with different reward functions exhibit varying performance. FR-LH and E-FR-LH are committed to achieve network long-term fairness and have comparable network capacity and reuse performance. However, the delivery ratio of these two methods are different. As shown in Fig. \ref{fig:noCrashCompare}(b), the delivery ratio of E-FR-LH is 23.82\% higher than that of FR-LH as $N$=4. We can conclude that as the number of nodes increases, E-FR-LH optimizes the delivery ratio more effectively. Figure \ref{fig:noCrashCompare} also confirms the validity of SECOPA when network deployment changes; training the power allocation model according to new topology and node configuration is all that is required.

\begin{figure}
\begin{center}
\includegraphics[width=3.3in]{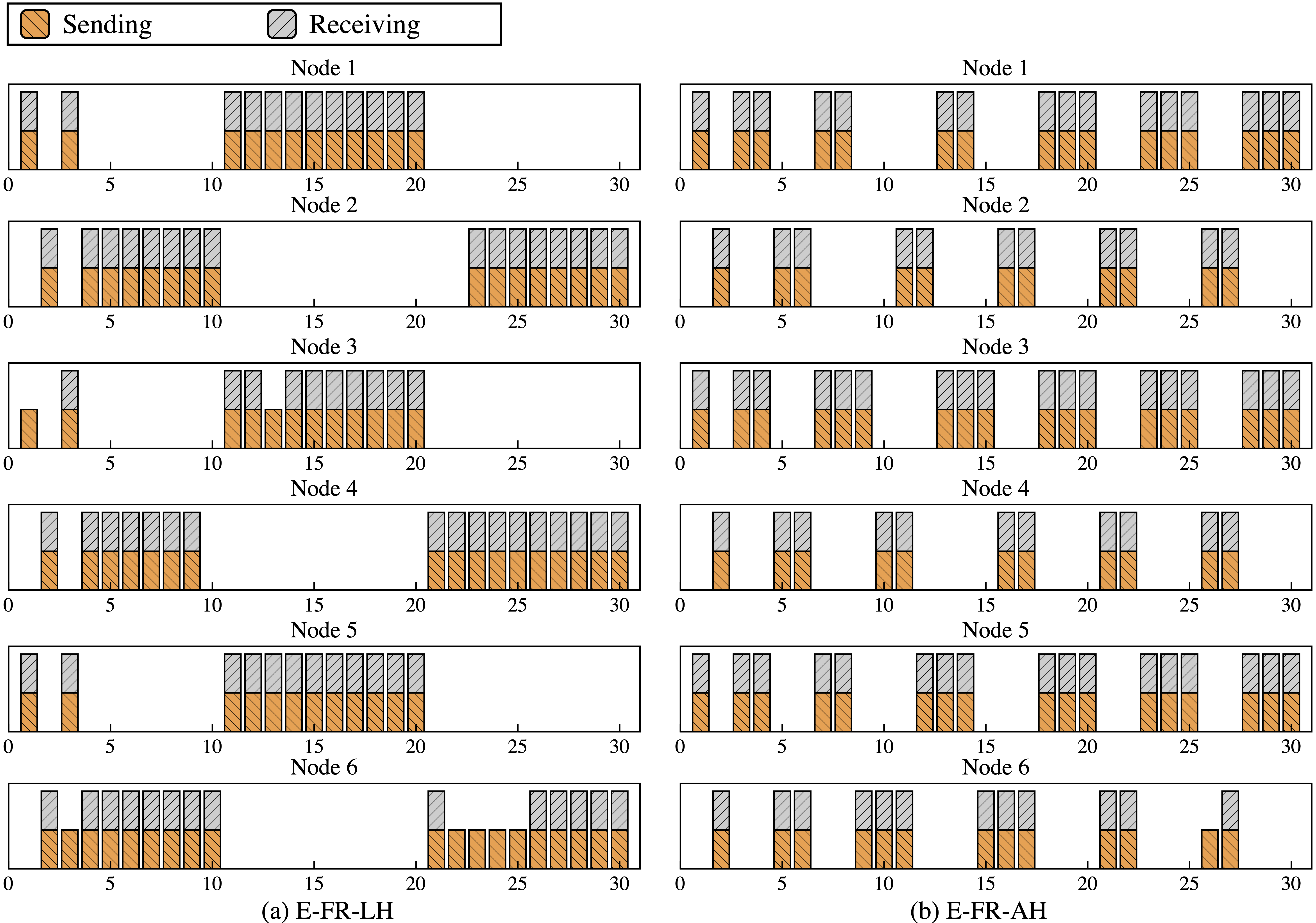}
\vspace{-1em}
\caption{Behaviors of (a) E-FR-LH, and (b) E-FR-AH.} 
\label{fig:E-FR-behavior2}
\end{center}
\end{figure}

E-FR-LH and E-FR-AH use lifetime-horizon fairness and adaptive-horizon fairness, respectively, for model evaluation throughout the training phase. Figure \ref{fig:noCrashCompare} shows that E-FR-AH outperforms E-FR-LH in all scenarios. To investigate the reasons for the differences between the two implementations of SECOPA (i.e., E-FR-LH and E-FR-AH), we further exhibit their sending and receiving behaviors in Fig. \ref{fig:E-FR-behavior2}. Node behavior is observed to be periodic in E-FR-LH and more continuous in E-FR-AH. E-FR-LH strives for long-term network fairness, where all nodes are expected to have the same amount of communication when network services are terminated. As shown in Fig. \ref{fig:E-FR-behavior2}(a), after a few consecutive transmissions, nodes withdraw from contention for communication resources to provide a better communication environment for other nodes and to reduce potential energy wastage from fail transmissions. E-FR-AH evaluates communication fairness over a shorter time horizon, and nodes are expected to communicate successfully in every time slot. As shown in Fig. \ref{fig:E-FR-behavior2}(b), nodes in E-FR-LH transmitted 88 times, of which 81 were successful, whereas nodes in E-FR-AH transmitted 86 times, of which 85 were successful.

\begin{table}[htp]
\caption{Comparisons of average delay as node density increases.}
\vspace{-1em}
\begin{center}
\begin{tabular}{lcccc}
\hline
Method&$N$=3&$N$=4&$N$=5&$N$=6\\
\hline
Greedy&0.20&1.09&15.22&N/A\\
Random&0.95&1.70&2.95&4.97\\
N-TDMA&4.45&6.34&8.35&10.02\\
CQL&0.88&17.46&25.79&8.18\\
DR-DLMA&0.17&1.05&2.00&1.37\\
\hline
SECOPA\textbf{[ours]}&&&&\\
FR-LH&0.14&0.78&1.84&1.25\\
E-FR-LH&0.14&0.82&1.62&1.29\\
E-FR-AH&\textbf{0.11}&\textbf{0.75}&\textbf{1.61}&\textbf{1.21}\\
\hline
\end{tabular}
\end{center}
\label{tab:nocrashdelay}
\end{table}

\begin{figure*}
\begin{center}
\includegraphics[width=6in]{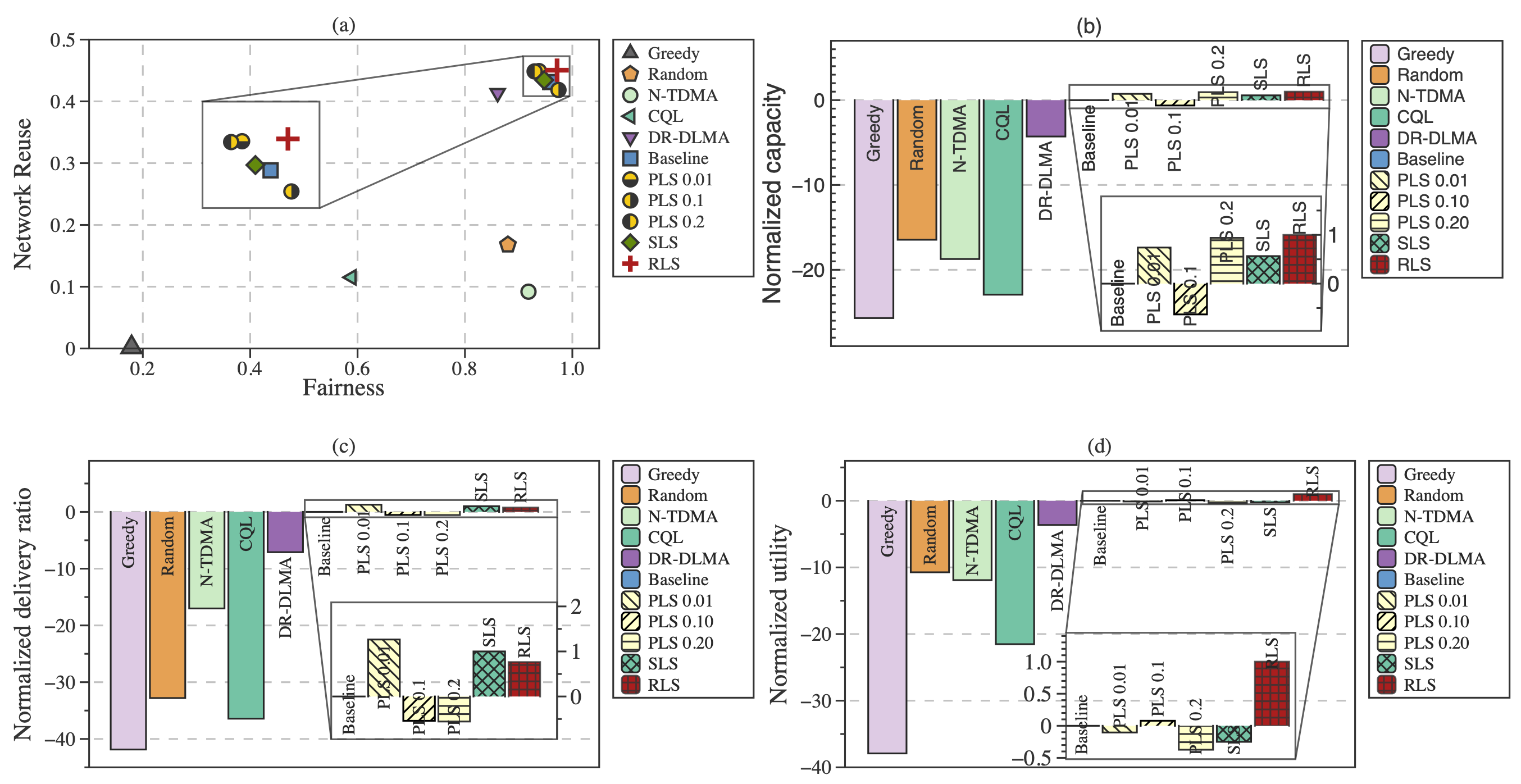}
\vspace{-1em}
\caption{Effects of baseline methods and SECOPA with different learning strategies on (a) fairness and reuse; (b) normalized capacity, (c) normalized delivery ratio; and (d) normalized network utility in an IC-UASN with six transmitter-receiver pairs, $\epsilon=0.01$.}
\label{fig:6all001}
\end{center}
\end{figure*}

\begin{figure*}
\begin{center}
\includegraphics[width=6in]{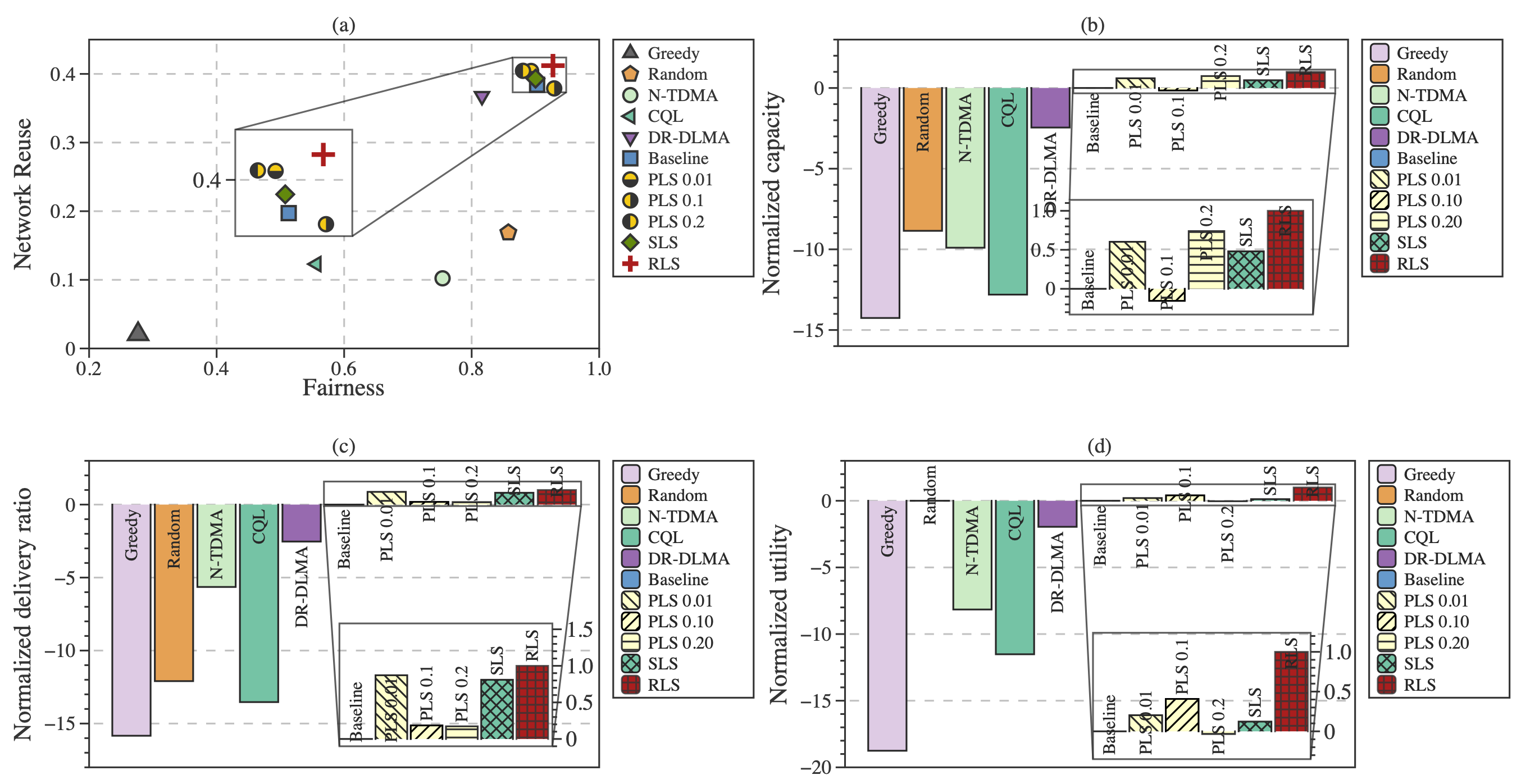}
\vspace{-1em}
\caption{Effects of baseline methods and SECOPA with different learning strategies on (a) fairness and reuse; (b) normalized capacity, (c) normalized delivery ratio; and (d) normalized network utility in an IC-UASN with six transmitter-receiver pairs, $\epsilon=0.1$.}
\label{fig:6all01}
\end{center}
\end{figure*}

\begin{figure*}
\begin{center}
\includegraphics[width=6in]{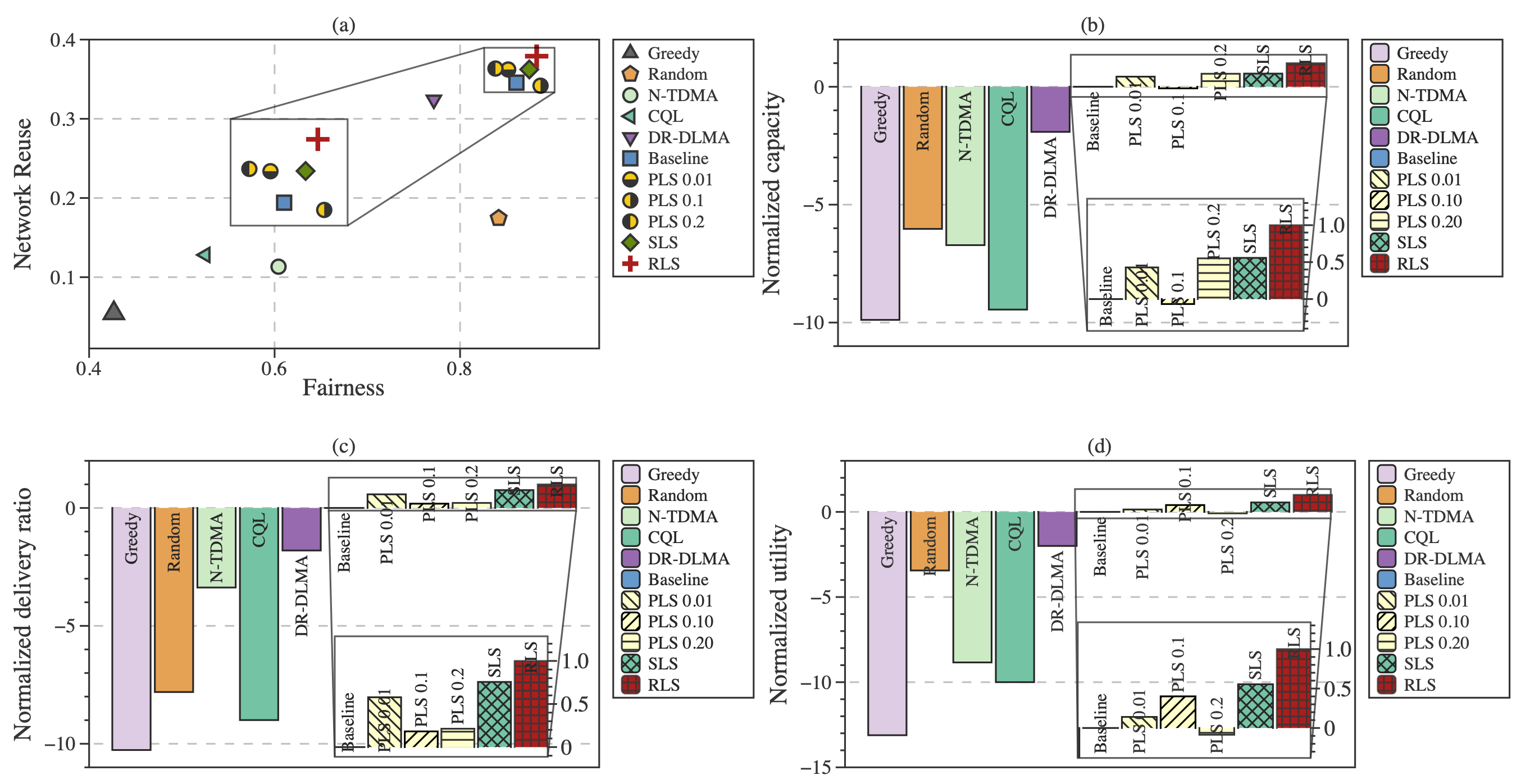}
\vspace{-1em}
\caption{Effects of baseline methods and SECOPA with different learning strategies on (a) fairness and reuse; (b) normalized capacity, (c) normalized delivery ratio; and (d) normalized network utility in an IC-UASN with six transmitter-receiver pairs, $\epsilon=0.2$.}
\label{fig:6all02}
\end{center}
\end{figure*}

Table \ref{tab:nocrashdelay} lists the average delivery delay of the different transmission strategies as the network density increases. Greedy failed the network when $N$=6, resulting in an unpredictable delivery delay. Random performs better than N-TDMA because it allows simultaneous communications, while N-TDMA has only one communication per time slot. However, we can extrapolate from the trend that N-TDMA will outperform Random as network density continues to increase. SECOPA (E-FR-AH) has relatively lower delivery delays, from 30.57\% (Greedy, $N$=4) up to 97.40\% (N-TDMA, $N$=3), than the baselines, demonstrating that our proposed scheme has advantages in time-sensitive applications. In addition, SECOPA with adaptive evaluation horizon always has lower delivery delays than FR-LH and E-FR-LH.

Based on the simulation results presented above, it can be concluded that SECOPA with E-FR-AH implementation achieves a better balance between individual QoS requirements and global fair-effective communication than other solutions. In the subsequent section, we use E-FR-AH to compare the robustness of different learning strategies.

\subsection{Results on Learning Strategies}

Among the baseline methods and SECOPA variants, the five baseline methods have no learning strategies, and all SECOPA variants are based on E-FR-AH standard. SECOPA is trained using three advanced learning strategies (i.e., PLS, SLS, RLS). PLS 0.01, PLS 0.10, and PLS 0.20 refer to PLS with $\epsilon$=0.01, $\epsilon$=0.1, and $\epsilon$=0.2, respectively. SLS has an incremental $\epsilon$ ranging from $[0,0.2]$, while RLS modifies $\epsilon$ responsive to model performance. All scenarios contain a mobile node and six transmitter-receiver pairs. Baseline refers specifically to the model implemented by E-FR-AH and trained in a UASN without node malfunctions. Figures \ref{fig:6all001}-\ref{fig:6all02} display the effects of learning strategies on network performance in IC-UASNs with $\epsilon$=0.01, $\epsilon$=0.1, and $\epsilon$=0.2, respectively.

For each baseline method and learning strategy, its normalized performance $\hat{x}$ is defined as $\hat{x}$=$\frac{x-x^{Baseline}}{max-x^{Baseline}}$, where $x$ is the achieved performance, $max$ is the maximum performance achieved by all learning strategies, and $x^{Baseline}$ is the performance achieved by Baseline. We use $x^{Baseline}$ as a normalized minimum to visually evaluate the effectiveness of learning strategies in imperfect environments. The normalized capacity $\hat{\mathcal{C}}$, normalized delivery ratio $\hat{\mathcal{D}}$, and normalized utility $\hat{\mathcal{U}_{\mathcal{N}}}$ are calculated as $\hat{\mathcal{C}_{\mathcal{N}}}$=$\frac{\mathcal{C}_{\mathcal{N}}-\mathcal{C}_{\mathcal{N}}^{Baseline}}{\mathcal{C}_{\mathcal{N}}^{max}-\mathcal{C}_{\mathcal{N}}^{Baseline}}$, $\hat{\mathcal{D}}$=$\frac{\mathcal{D}-\mathcal{D}^{Baseline}}{\mathcal{D}^{max}-\mathcal{D}^{Baseline}}$, and $\hat{\mathcal{U}_{\mathcal{N}}}$=$\frac{\mathcal{U}_{\mathcal{N}}-\mathcal{U}_{\mathcal{N}}^{Baseline}}{\mathcal{U}_{\mathcal{N}}^{max}-\mathcal{U}_{\mathcal{N}}^{Baseline}}$, respectively.

Figure \ref{fig:6all001} demonstrates the effects of learning strategies on fairness and reuse, normalized capacity, normalized delivery ratio, and normalized utility when $\epsilon$=$0.01$. The SECOPA variants have much higher network reuse when compared with the baseline methods, as Fig. \ref{fig:6all001}(a) shows. There are small differences between the variants. PLS 0.1 provides advantages in terms of communication fairness, but is not effective in increasing network reuse, and thus has lower network capacity and delivery ratio, as shown in Fig. \ref{fig:6all001}(b) and (c). PLS 0.01 and PLS 0.2 achieve similar network reuse as RLS, while they have lower fairness than other SECOPA variants. The network reuse achieved by N-TDMA, Random, and CQL is far below that of our schemes. The normalized capacity and delivery ratio are shown in Figure \ref{fig:6all001}(b) and (c), where RLS performs best. Figure \ref{fig:6all001}(d) shows that RLS and PLS 0.1 provide greater network utility than Baseline, while the other methods may have difficulty adapting to the imperfect environments as expected.

Figure \ref{fig:6all01} illustrates the effects of learning strategies on various network performance when $\epsilon$=$0.1$. Baseline shows a significant decrease in network reuse when $\epsilon$ is changed from 0.01 to 0.1. RLS and PLS 0.1 have better communication fairness than other methods, while RLS dominates in network reuse. Baseline, the three PLS methods, and Random share the same tendencies as seen in Fig. \ref{fig:6all001}(a), but N-TDMA and Greedy are different. N-TDMA degrades network reuse and communication fairness, while Greedy improves both network reuse and fairness. Figures \ref{fig:6all01}(b)-(d) show that RLS has superior network capacity, delivery ratio, and network utility than other schemes.

Figure \ref{fig:6all02} shows the effects of learning strategies on network performance when $\epsilon$=$0.2$. As $\epsilon$ continues to increase, all but Greedy experience performance degradation. Compared to Fig. \ref{fig:6all001}(a) and Fig. \ref{fig:6all01}(a), N-TDMA suffers an obvious degradation in fairness (by 34.2\% and 19.9\%, respectively), while Greedy achieves greater network reuse. Figures \ref{fig:6all02}(b)-(d) show the superiority of RLS over other schemes in terms of network capacity, delivery ratio, and network utility.

Based on the above experimental results, several important observations can be deduced. \textit{First}, as $\epsilon$ increases, the fairness of N-TDMA experiences a significant decline. In contrast, Greedy exhibits improvements in both network reuse and communication fairness. This makes sense because as more nodes resort to random transmission or refrain from competing for the shared acoustic channel, the number of concurrent communications decreases, leading to enhanced communication performance under the Greedy scheme. In the case of N-TDMA, a higher malfunction rate increases the likelihood of concurrent communication during certain time slots, resulting in a slight rise in network reuse. However, regular N-TDMA nodes transmit less than random nodes, which diminishes communication fairness.

\textit{Second}, although CQL and DR-DLMA are also RL-based methods, they do not optimize network performance as originally anticipated. Neither method takes communication fairness into account, indicating that the cooperative optimization of network performance requires the consideration for communication fairness. \textit{Third}, by treating Baseline as a special case of PLS when $\epsilon$=0, we can discover that models trained with a fixed malfunction rate based on past experience do not necessarily perform better in similar environments, and that models trained from a zero-malfunction environment (RLS, SLS, and Baseline) have superior network utility to other models as $\epsilon$ increases. \textit{Finally}, despite both SLS and RLS-trained agents encountering varying malfunction rates during the training phase, RLS exhibits a stronger proficiency in handling these environments. Consequently, it is advisable to initially train the model in fully-cooperative environments and subsequently employ suitable advanced learning strategies to teach the model how to handle imperfect environments by maintaining semi-cooperative awareness.

\subsection{Results on Hyper-parameters}

\begin{figure}
\begin{center}
\includegraphics[width=2.9in]{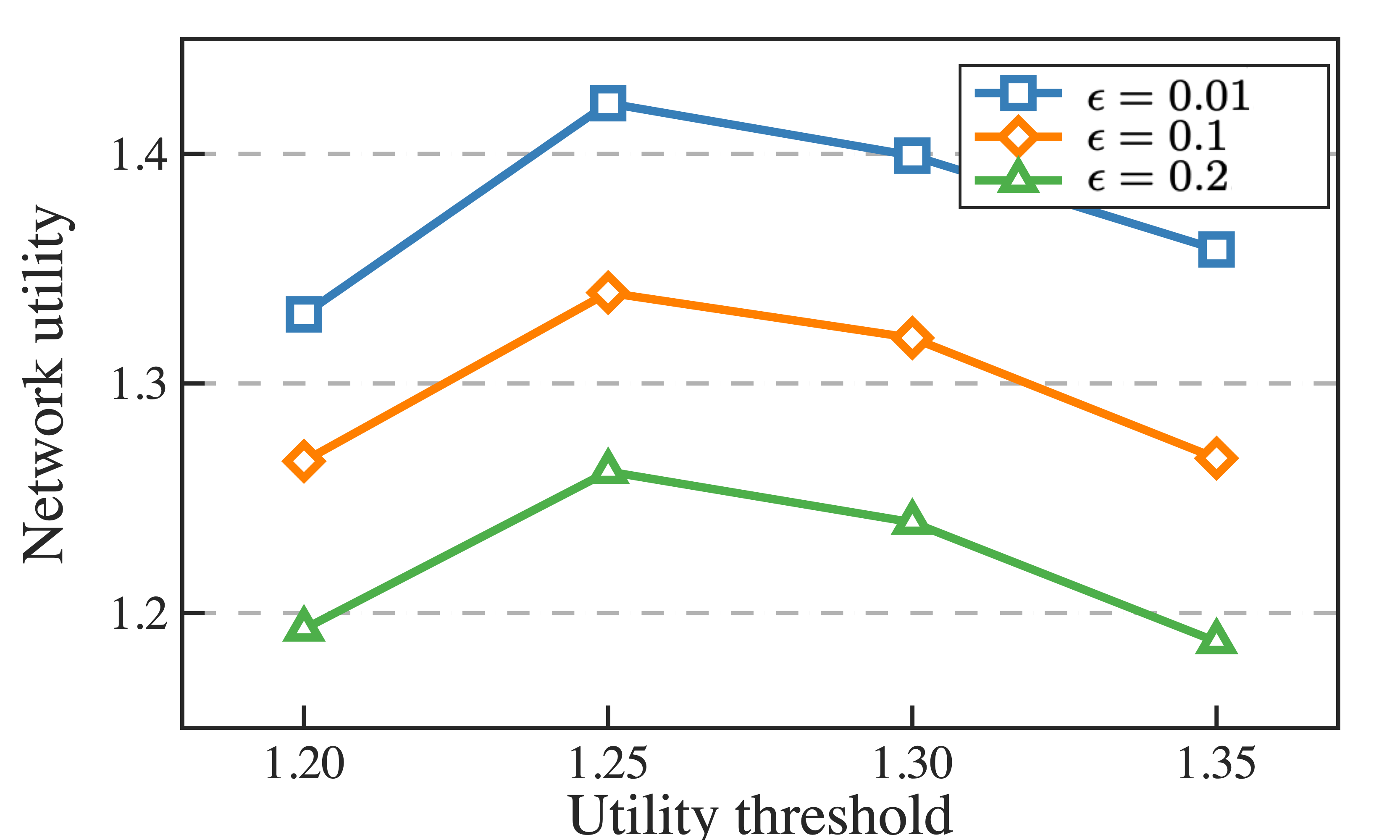}
\vspace{-1em}
\caption{Effects of utility threshold on network utility in IC-UASNs.}
\label{fig:Thd}
\end{center}
\end{figure}

\begin{figure}
\begin{center}
\includegraphics[width=2.9in]{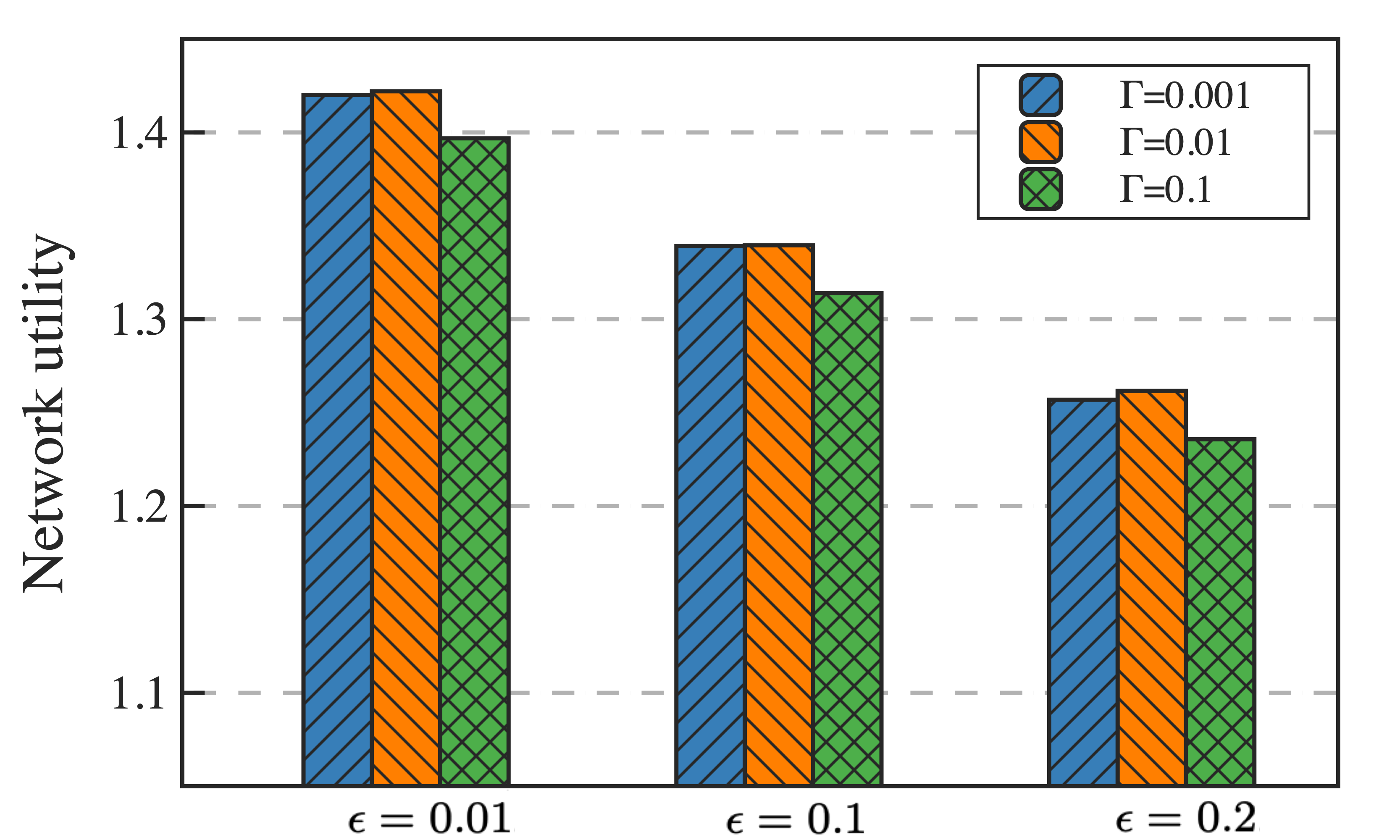}
\vspace{-1em}
\caption{Effects of advanced learning factor on utility performance in IC-UASNs.} 
\label{fig:learningRate}
\end{center}
\end{figure}

Figures \ref{fig:Thd} and \ref{fig:learningRate} illustrate the impact of utility threshold ($v_{b}^{0}$) and advanced learning factor ($\Gamma$) on network utility in imperfect UASNs, respectively. All scenarios include a mobile node and six transmitter-receiver pairs. It can be observed from Fig. \ref{fig:Thd} that the network utility increases as $v_{b}^{0}$ increases from 1.20 to 1.25, and then tends to decrease. In all scenarios, the maximum network utility is achieved when $v_{b}^{0}$ reaches 1.25, and the achieved performance decreases as the node malfunction rate increases. By comparing the achieved network utility with $v_{b}^{0}$, we evaluate whether the model has adapted to the current environment. Suppose $\mathcal{U}_{\mathcal{N}}\!>\!v_{b}^{0}$, the model is considered capable of handling current environments, and the IE-Generator increases the malfunction rate; otherwise, the generator decreases the malfunction rate to reduce the training difficulty. Consequently, increasing $v_{b}^{0}$ forces the model to become more transparent to the current environment. However, as shown in Fig. \ref{fig:Thd}, extensive $v_{b}^{0}$ can cause the model to fail training. Since the performance of the model is unlikely to exceed the threshold at each evaluation epoch, the IE-Generator will continuously reduce the malfunction rate. Bounded by $[0,\epsilon^{u}]$, the model is trained in situations with a malfunction rate close to zero, resulting in a performance comparable to Baseline. It can be observed from Fig. \ref{fig:Thd} that models trained with $v_{b}^{0}$=$1.25$ have a higher network utility than models trained with other settings.

We then evaluate the effect of the advanced learning factor $\Gamma$ in IC-UASNs, which controls how much the IE-Generator  adjust the malfunction rate in response to model performance. A small $\Gamma$ instructs the IE-Generator to provide similar environments for adjacent training episodes, resulting in a stable training process. Conversely, a large $\Gamma$ may lead to unstable training process. The achieved network utility over different malfunction rate $\epsilon$ and advanced learning factor $\Gamma$ is shown in Fig. \ref{fig:learningRate}. With the increase of $\epsilon$, the network utility of all models tends to decrease, as more nodes degrade to normal nodes that perform random actions or stop transmitting. We can also observe that models with $\Gamma=0.01$ always achieve better performance than those with larger or smaller $\Gamma$. This observation may indicate that a moderate learning factor (which seeks a balance between training stability and diversity) is beneficial for the model to handle imperfect networks.

\section{Conclusion}\label{sec:con}

In this paper, we have formulated the fair-effective communication and robustness optimization problem in imperfect and energy-constrained UASNs (IC-UASNs). The main challenge of this problem is that the intelligent nodes are energy constrained and may not be able to load the cooperative optimization algorithms or access the input information in some time slots. Specifically, we have proposed SECOPA, a semi-cooperative approach to guide the power allocation tasks for achieving fair-effective communication and robustness in IC-UASNs. SECOPA narrows the gap between simulation and reality by 1) constructing a relatively comprehensive simulator to characterize the acoustic communication propagation process, 2) optimizing network performance under the constraint of limited node energy supply and network lifetime, and 3) devising learning algorithms to generate imperfect training environments for underwater node malfunctions. Numerical results have demonstrated the superiority of our proposed SECOPA in achieving fair-effective communication and robustness in IC-UASNs. Guidance for constructing a semi-cooperative optimization model in an imperfect environment is also provided. We expect that SECOPA will maximize the performance of UASNs in real-world conditions and enable a broader range of underwater applications such as SAGAIN.

There are a few limitations that will be explored in our future work. First, the utility threshold is currently predetermined. A learned or adaptive threshold is left for future research. Second, we discovered that while using a smaller advanced learning factor yields better performance, it may also prolong the training process. The relationship between advanced learning factor and training efficiency needs further investigation. Third, this research primarily focuses on the software-related malfunctions. A more general model that considers other possible malfunction scenarios will also be our future work. Finally, time synchronization is important for UASNs. For future work, we will incorporate a time synchronization scheme into the performance optimization strategy of UASNs, thus enabling a wider range of underwater applications.

 \section*{Acknowledgments}
This work was supported by the National Key Research and Development Program (Grant No. 2021YFC2803000, SQ2020YFB050001), the National Natural Science Foundation (Grants No. 6197011140, No. 62101211, and No. 61971206).

\ifCLASSOPTIONcaptionsoff
  \newpage
\fi



%
\bibliographystyle{IEEEtran}
\bibliography{IEEEabrv,mybibfile}

%

\begin{IEEEbiography}[{\includegraphics[width=1in,height=1.25in,clip,keepaspectratio]{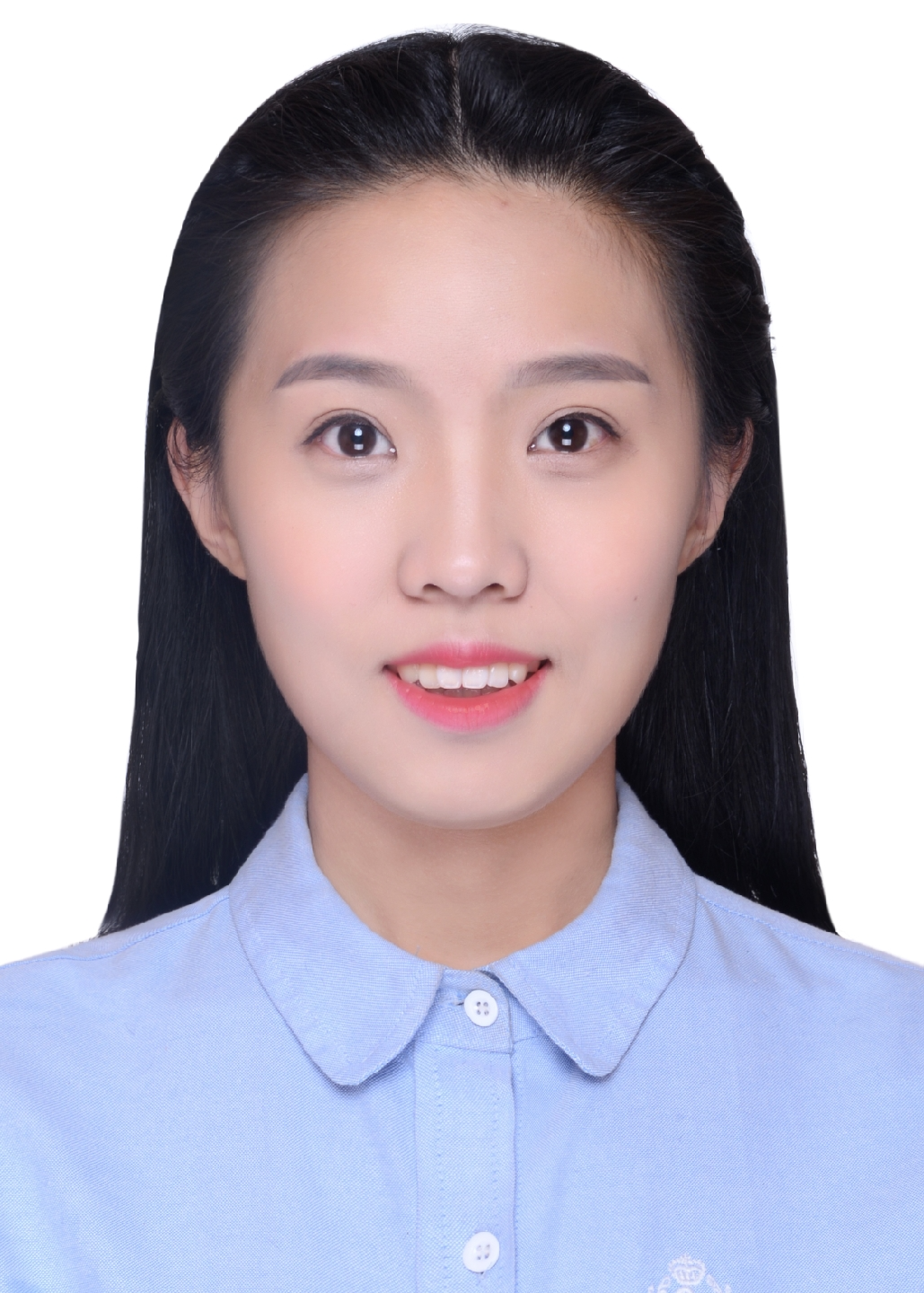}}]{Yu Gou} received her B.S. degree in 2015 and M.S. degree in 2018 from the College of Computer Science and Technology, Jilin University, Changchun, China, where she is currently pursuing the Ph.D. degree. Recently, her research interests are include deep multi-agent reinforcement learning and underwater network performance optimization.
\end{IEEEbiography}
\begin{IEEEbiography}[{\includegraphics[width=1in,height=1.25in,clip,keepaspectratio]{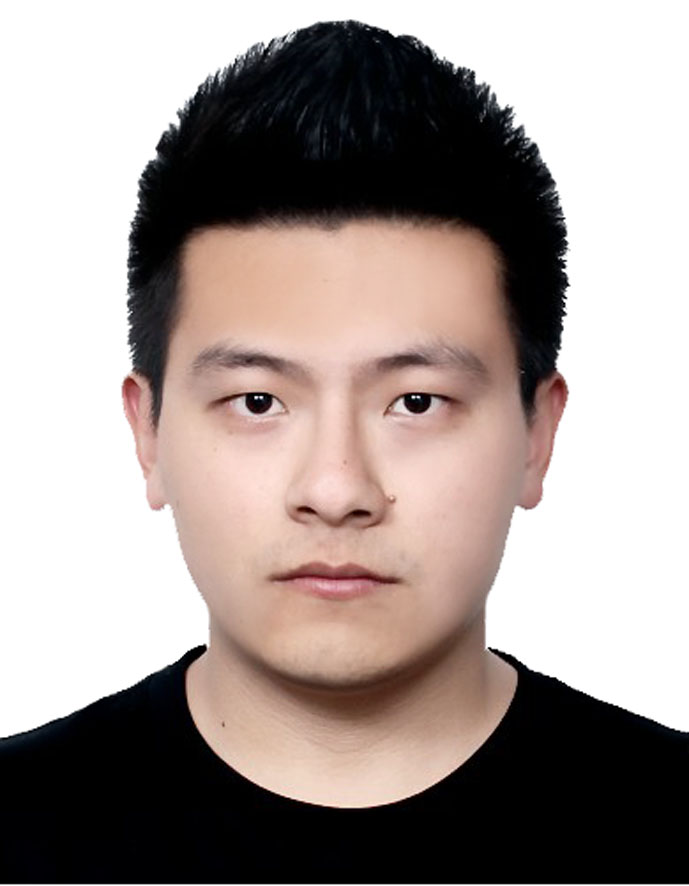}}]{Tong Zhang} received his B.S. degree in 2014 from the College of Mathematics and Computer Science, Fuzhou University, Fuzhou, China and M.S. degree in 2018 from the College of Computer Science and Technology, Jilin University, Changchun, China, where he is currently pursuing the Ph.D. degree. His research interests are mainly focus on the resource management of underwater wireless sensor networks and deep reinforcement learning.
\end{IEEEbiography}
\begin{IEEEbiography}[{\includegraphics[width=1in,height=1.25in,clip,keepaspectratio]{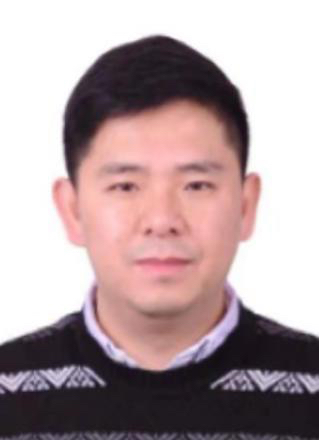}}]{Jun Liu} received the BEng degree (2002) in computer science from Wuhan University, China, the PhD degree (2013) in Computer Science and Engineering from University of Connecticut, USA. Currently, he is a professor of the School of Electronic and Information Engineering at Beihang University, China. His major research focuses on underwater networking, synchronization, and localization.
\end{IEEEbiography}
\begin{IEEEbiography}[{\includegraphics[width=1in,height=1.25in,clip,keepaspectratio]{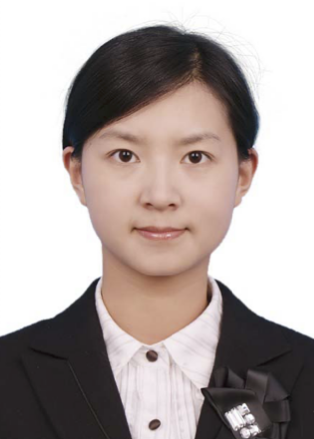}}]{Tingting Yang} (M’13) received her B.Sc. (2004) and Ph.D. degrees (2010) from Dalian Maritime University, China. She is currently a Professor at Dalian Maritime University, and also at Pengcheng Laboratory. Her research interests are in the areas of space-air-ground-sea integrated networks, and network AI. She serves as the associate Editor-in-Chief of the IET Communications, as well as the advisory editor for SpringerPlus.
\end{IEEEbiography}
\begin{IEEEbiography}[{\includegraphics[width=1in,height=1.25in,clip,keepaspectratio]{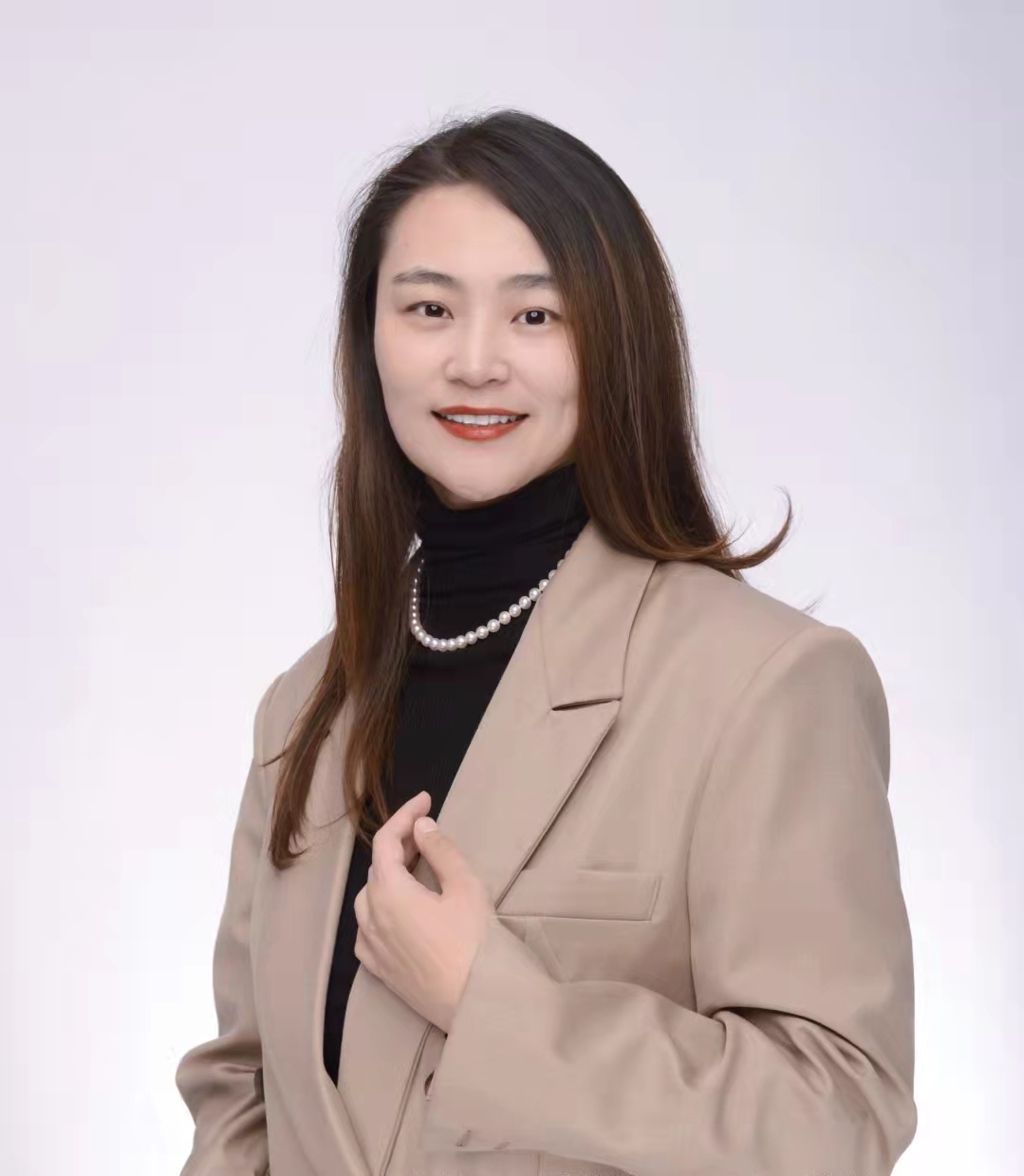}}]{Shanshan Song} received the BS degree (2011) and MS degree (2014) in computer science and technology from Jilin University, China, received PhD degree (2018) in Management science and engineering from Jilin University, China. Currently, she is the associate professor of the College of Computer Science and Technology at Jilin University, China. Her major research focuses on underwater  localization and navigation and machine learning.
\end{IEEEbiography}
\begin{IEEEbiography}[{\includegraphics[width=1in,height=1.25in,clip,keepaspectratio]{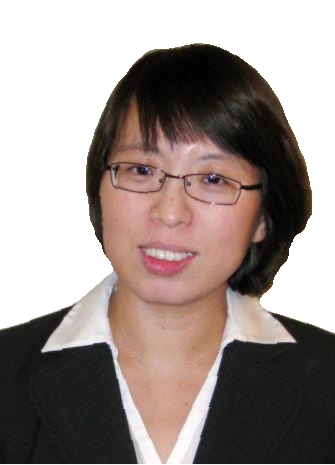}}]{Jun-Hong Cui} received the BS degree (1995) in computer science from Jilin University, China, the MS degree (1998) in computer engineering from Chinese Academy of Sciences, and the PhD degree (2003) in computer science from UCLA. Currently, she is on the faculty of the College of Computer Science and Technology, Jilin University, China. Recently, her research mainly focuses on exploiting the spatial properties in the modeling of network topology, algorithm and protocol design in underwater sensor networks.
\end{IEEEbiography}




\end{document}